\documentclass[aps,prd,preprintnumbers,groupedaddress,nofootinbib,amssymb,eqsecnum,epsfig]{revtex4}
\usepackage{graphicx}
\usepackage{bm}
\usepackage{amsmath}
\usepackage{color}
\usepackage{amsfonts}

\begin{document}
\newcommand{\newc}{\newcommand}

\newcommand{\ben}{\begin{eqnarray}}
\newcommand{\een}{\end{eqnarray}}
\newc{\be}{\begin{equation}}
\newc{\ee}{\end{equation}}
\newc{\ba}{\begin{eqnarray}}
\newc{\ea}{\end{eqnarray}}
\newc{\bea}{\begin{eqnarray*}}
\newc{\eea}{\end{eqnarray*}}
\newc{\D}{\partial}
\newc{\ie}{{\it i.e.} }
\newc{\eg}{{\it e.g.} }
\newc{\etc}{{\it etc.} }
\newc{\etal}{{\it et al.}}
\newcommand{\nn}{\nonumber}
\newc{\ra}{\rightarrow}
\newc{\lra}{\leftrightarrow}
\newc{\lsim}{\buildrel{<}\over{\sim}}
\newc{\gsim}{\buildrel{>}\over{\sim}}

\title{Observational constraints on generalized Proca theories} 

\author{
Antonio De Felice$^{1}$, 
Lavinia Heisenberg$^{2}$, and 
Shinji Tsujikawa$^{3}$}

\affiliation{
$^1$Center for Gravitational Physics, Yukawa Institute for Theoretical Physics,\\
Kyoto University, 606-8502, Kyoto, Japan
$^2$Institute for Theoretical Studies, ETH Zurich, Clausiusstrasse 47, 8092 Zurich, Switzerland\\
$^3$Department of Physics, Faculty of Science, Tokyo University of Science, 1-3, Kagurazaka,
Shinjuku-ku, Tokyo 162-8601, Japan}

\date{\today}

\begin{abstract}

In a model of the late-time cosmic acceleration within 
the framework of generalized Proca theories, there exists 
a de Sitter attractor preceded by the dark energy equation of state 
$w_{\rm DE}=-1-s$, where $s$ is a positive constant.
We run the Markov-Chain-Monte-Carlo code to confront the model 
with the observational data of Cosmic Microwave
Background (CMB), baryon acoustic oscillations, 
supernovae type Ia, and local measurements of the Hubble 
expansion rate for the background cosmological solutions 
and obtain the bound $s=0.254^{{}+ 0.118}_{{}-0.097}$ 
at 95\,\% 
confidence level (CL). Existence of the additional parameter $s$ 
to those in the $\Lambda$-Cold-Dark-Matter ($\Lambda$CDM) 
model allows to reduce tensions of the Hubble constant 
$H_0$ between the CMB and the low-redshift 
measurements. Including the cosmic growth data of redshift-space distortions 
in the galaxy power spectrum and taking into account no-ghost 
and stability conditions of cosmological perturbations, we find that 
the bound on $s$ is shifted to $s=0.16^{+0.08}_{-0.08}$ (95\,\% CL)
and hence the model with $s>0$ is still favored over
the $\Lambda$CDM model. Apart from the quantities $s, H_0$ 
and the today's matter density parameter $\Omega_{m0}$,
the constraints on other model 
parameters associated with perturbations are less stringent, 
reflecting the fact that there are different sets of parameters 
that give rise to a similar cosmic expansion and growth history.

\end{abstract}


\preprint{YITP-17-58}

\maketitle

\section{Introduction}

Two fundamental pillars are used in the standard model of Big Bang
cosmology for describing the physics on cosmological scales: the
cosmological principle and General Relativity (GR). The first is the
notion of homogeneity and isotropy. Even if the fundamental theory
behind GR is very elegant and simple, the problems of cosmological
constant and dark energy imply that we may need some modifications of
GR in both infrared and ultraviolet scales.  The cosmological constant
problem represents the enormous discrepancy between observations and
the expectations from a field theory point of view \cite{Weinberg},
whereas the dark energy problem stands for the observed late-time
acceleration of the Universe. Another tenacious challenge is the
successful construction of a consistent theory of quantum gravity.  To
address such problems, there have been numerous attempts for modifying
gravity in the infrared and ultraviolet regimes
\cite{Copeland:2006wr,Sotiriou:2008rp,DeFelice:2010aj,Clifton:2011jh,Bull:2015stt,
  Amendola:2012ys,Joyce:2014kja,Amendola:2016saw}.

In the context of infrared modifications of gravity, theories based on 
scalar fields are the most extensively explored ones. One essential reason for these considerations is the natural provision of isotropic accelerated expansion. Besides this practical reasoning, we know that scalar fields do exist in nature. The Higgs field is a fundamental ingredient of the Standard Model of particle physics. Accepting an additional scalar degree of freedom in the gravity sector, the scalar field has to be very light to drive the late-time cosmic acceleration. 
This new scalar degree of freedom generally gives rise to
long-range forces with baryonic matter, but such fifth forces have never been detected in solar-system tests of gravity \cite{Will}. 
Therefore, one has to rely on some successful implementations 
of screening mechanisms, which hide the scalar field on small scales whereas being unleashed on large scales to produce desired cosmological effects. In particular, the Vainshtein mechanism \cite{Vainshtein}
in the presence of non-linear field self-interactions can efficiently screen the propagation of fifth forces within a radius much larger than the solar-system 
scale \cite{Deffayet:2001pu,Babichev:2009jt,Burrage:2010rs,DeFelice:2011th,Kimura:2011dc,Kase:2013uja}.

An interesting class of the self-interacting scalar field with a Galileon symmetry 
was proposed in Ref.~\cite{Nicolis:2008in}. 
These Galileon interactions involve explicit dependence on second derivatives of the scalar field in their Lagrangians, but they maintain the second-order nature of field equations such that the Ostrogradski instability is avoided. 
A naive covariantization of these Galileon interactions would result in higher-order equations of motion. This can be prevented by including corresponding non-minimal derivative couplings 
with the Ricci scalar and the Einstein tensor \cite{cova1,cova2}. 
The generalization of covariant Galileons led to the construction of theories respecting the Galilean symmetry on the de Sitter background \cite{Burrage:2011bt} 
and the rediscovery of the Horndeski action \cite{Horndeski}. 

Horndeski theories constitute the most general scalar-tensor theories leading to 
second-order equations of motion with one scalar propagating degree of freedom 
besides two tensor polarizations \cite{Horndeski,Deffayet:2011gz,Kobayashi:2011nu,Charmousis:2011bf}. Similar scalar-tensor theories with second-order equations of motion also arise from the decoupling limit of massive gravity \cite{deRham:2010tw,deRham:2011by,Burrage:2011cr,Heisenberg:2014kea}. 
Even outside the second-order domain, it is possible to construct more general 
scalar-tensor theories with one scalar propagating degree of 
freedom \cite{Zumalacarregui:2013pma,Gleyzes:2014dya,Langlois:2015cwa}. 
All these new realms of possibilities have been giving rise to a plethora of attempts for describing dark energy. 
These attempts also shed light to how classical field theories can be constructed in a consistent way as to keep the theory sensible and viable, i.e., without introducing ghost instabilities and removing unwanted degrees 
of freedom.

The Standard Model of particle physics contains both abelian and non-abelian 
vector fields as the fundamental carriers of  gauge interactions. 
Consequently, it is comprehensible to wonder whether bosonic vector fields 
may also play an important role in the cosmological evolution besides scalar fields. Similarly to the scalar counterpart, vector fields being part of the gravitational interactions could naturally result in an accelerated Universe on large scales while being screened on small scales. Learned from the lessons for constructing consistent theories for scalar-tensor interactions, one can apply the same approach to vector-tensor theories. In Minkowski space-time, allowing for a mass of 
the vector field leads to the propagation of a longitudinal scalar mode besides two transverse vector polarizations due to the breaking of $U(1)$ gauge invariance. One can generalize this massive Proca theory to that in curved 
space-time in such a way that the propagating degrees of freedom 
remain three besides two tensor polarizations. A specific type of massive Proca theories naturally arise in the framework of Weyl geometries \cite{Jimenez:2015fva,Jimenez:2016opp}.

The generalized Proca theories with second-order equations of motion constitute Galileon type vector self-interactions with non-minimal derivative couplings to gravity. The systematic construction of the action of generalized Proca theories was carried out in Ref.~\cite{Heisenberg:2014rta}, where it was shown that despite the derivative self-interactions only three physical degrees of freedom propagate. 
The specific case where the longitudinal mode of the vector field has the scalar Galilean self-interactions was considered in Ref.~\cite{Tasinato:2014eka}. These generalized Proca interactions were further investigated in Refs.~\cite{Allys:2015sht,Jimenez:2016isa}. Even outside the domain of 
second-order theories, one can construct more general vector-tensor 
interactions without increasing the number of propagating degrees of 
freedom relative to that in generalized Proca theories \cite{Heisenberg:2016eld,Kimura:2016rzw}.  

Recently, the background cosmological solutions and the stabilities of 
perturbations were studied for concrete dark energy models that belong to 
generalized Proca theories \cite{DeFelice:2016yws,DeFelice:2016uil}. 
These models can be also compatible with solar-system constraints for 
a wide range of parameter space \cite{DeFelice:2016cri}. 
At the background level, there exists a de Sitter attractor responsible 
for the late-time cosmic acceleration. 
Moreover, it was illustrated how dark energy models in the framework of generalized Proca theories can be observationally distinguished from the standard cosmological model according to both expansion history and cosmic growth \cite{DeFelice:2016yws, DeFelice:2016uil}. 
Within this framework of generalized Proca interactions, 
the de Sitter solution arises from the temporal component of the vector field compatible with the symmetries of homogeneity and isotropy 
of the Friedmann-Lema\^{i}tre-Robertson-Walker (FLRW) background. 
Even if the temporal component does not correspond to a propagating degree of freedom by construction, it does have a non-trivial effect on the cosmological dynamics by behaving as an auxiliary field. 
Another way of constructing homogeneous and isotropic cosmological 
solutions within this class of theories consists of considering triads configuration \cite{Jimenez:2016upj, Emami:2016ldl}. A third possibility is a combination of the temporal configuration with the triads as proposed in Ref.~\cite{Jimenez:2016upj}. It would be worthwhile to investigate the cosmological implications of this type of multi-Proca interactions and extend the existing studies \cite{Hull:2014bga, Jimenez:2013qsa, BeltranJimenez:2017cbn, Hull:2015uwa, Allys:2016kbq, Jimenez:2016upj}.

In this work, we go along the lines of  Refs.~\cite{DeFelice:2016yws, DeFelice:2016uil} and place constraints on these models by using several different cosmological data sets: the Cosmic Microwave Background (CMB) shift parameters 
(from the Planck data), 
Baryon Acoustic Oscillations (BAO), Supernova Type Ia (SN Ia) 
standard candles (from the Union 2.1 data), local measurements of the Hubble 
expansion rate, and Redshift-Space-Distortions (RSD).
The $\Lambda$-Cold-Dark-Matter ($\Lambda$CDM) model requires typical values of today's density parameter of non-relativistic matter $\Omega_{m0}$ to be 
around 0.31, whereas, in generalized Proca theories, 
smaller values of $\Omega_{m0}$ are reached due to the existence 
of an extra parameter, $s$. Moreover, the Hubble 
constant $H_0$ tends to be higher in generalized Proca theories 
with a lower total $\chi^2$ relative to that in the $\Lambda$CDM model.
Therefore, generalized Proca theories can help to reduce the tension 
between early-time and late-time data sets, compared to the 
$\Lambda$CDM model.

This paper is organized as follows.
In Sec.~\ref{sec:GenProca} we review the background cosmological 
dynamics for a class of dark energy models in the framework of 
generalized Proca theories.
In Sec.~\ref{persec} we discuss stability conditions and the evolution 
of matter density perturbations relevant to the growth 
of large-scale structures.
In Sec.~\ref{datasec} we explain the data sets used for the 
likelihood analysis in later sections.
In Sec.~\ref{backconsec} we place observational bounds on model 
parameters associated with the background expansion history 
by using the data of CMB, BAO, SN Ia, and Hubble expansion rate.
In Sec.~\ref{perconsec} we put constraints on model parameters related to 
linear cosmological perturbations by adding the RSD data in the analysis.
Sec.~\ref{consec} is devoted to conclusions.

\section{Generalized Proca theories and the background dynamics} 
\label{sec:GenProca}

In generalized Proca theories, the vector field $A^{\mu}$ possesses two transverse polarizations and one longitudinal scalar mode non-minimally coupled to gravity. To keep the equations of motion up to second order, 
the field self-interactions need to be of specific forms.
The action of generalized Proca theories is given by 
\cite{Heisenberg:2014rta, Jimenez:2016isa}
\begin{equation}
S=\int d^4x \sqrt{-g} \left( {\cal L}
+{\cal L}_M \right)\,,\qquad
{\cal L}=\sum_{i=2}^{6} {\cal L}_i\,,
\label{LagProca}
\end{equation}
where $g$ is the determinant of the metric tensor $g_{\mu \nu}$, 
${\cal L}_M$ is the matter Lagrangian density, and ${\cal L}_{2,3,4,5,6}$
are the vector-tensor interactions given by 
\begin{eqnarray}
{\cal L}_2 &=& G_2(X,F,Y)\,,
\label{L2}\\
{\cal L}_3 &=& G_3(X) \nabla_{\mu}A^{\mu}\,,
\label{L3}\\
{\cal L}_4 &=&
G_4(X)R+
G_{4,X}(X) \left[ (\nabla_{\mu} A^{\mu})^2
-\nabla_{\rho}A_{\sigma}
\nabla^{\sigma}A^{\rho} \right]\,,\label{L4} \\
{\cal L}_5 &=&
G_{5}(X) G_{\mu \nu} \nabla^{\mu} A^{\nu}
-\frac16 G_{5,X}(X) [ (\nabla_{\mu} A^{\mu})^3
-3\nabla_{\mu} A^{\mu}
\nabla_{\rho}A_{\sigma} \nabla^{\sigma}A^{\rho}
+2\nabla_{\rho}A_{\sigma} \nabla^{\gamma}
A^{\rho} \nabla^{\sigma}A_{\gamma}] \nonumber \\
& &-g_5(X) \tilde{F}^{\alpha \mu}
{\tilde{F^{\beta}}}_{\mu} \nabla_{\alpha} A_{\beta}\,,
\label{L5}\\
{\cal L}_6 &=& G_6(X) L^{\mu \nu \alpha \beta}
\nabla_{\mu}A_{\nu} \nabla_{\alpha}A_{\beta}
+\frac12 G_{6,X}(X) \tilde{F}^{\alpha \beta} \tilde{F}^{\mu \nu}
\nabla_{\alpha}A_{\mu} \nabla_{\beta}A_{\nu}\,.
\label{L6}
\end{eqnarray}
The field strength is $F_{\mu \nu}=\nabla_{\mu}A_{\nu}-\nabla_{\nu}A_{\mu}$ and its dual is $\tilde{F}^{\mu \nu}=\epsilon^{\mu \nu \alpha \beta}F_{\alpha \beta}/2$, where $\nabla_{\mu}$ stands for the covariant derivative operator.
The function $G_2$ can depend in general on the quantities 
\be
X =-\frac12 A_{\mu} A^{\mu}\,,\qquad
F = -\frac14 F_{\mu \nu} F^{\mu \nu}\,,\qquad 
Y = A^{\mu}A^{\nu} {F_{\mu}}^{\alpha}F_{\nu \alpha}\,,
\ee
while the remaining functions $G_{3,4,5,6}$ and $g_5$ depend only on $X$. 
The partial derivatives of the functions are denoted by $G_{i,X} \equiv \partial G_{i}/\partial X$. 
In the same way as in scalar Horndeski theories, the vector field is coupled only to the divergenceless tensors and their corresponding versions at the level of the equations of motion. Hence, the vector field is directly coupled to the Ricci scalar $R$, the Einstein tensor $G_{\mu \nu}=R_{\mu\nu}-R g_{\mu\nu}/2$, and the double dual Riemann tensor 
\begin{equation}
L^{\mu \nu \alpha \beta}=\frac14 \epsilon^{\mu \nu \rho \sigma} \epsilon^{\alpha \beta \gamma \delta} R_{\rho \sigma \gamma \delta}\,,
\end{equation}
where $\epsilon^{\mu \nu \rho \sigma}$ is the Levi-Civita tensor and 
$R_{\rho \sigma \gamma \delta}$ is the Riemann tensor.
The specific case with $G_2=m^2 X$ and $G_{3,4,5,6}=0$ corresponds to the standard Proca theory, in which case two transverse vector modes and the longitudinal scalar propagate. 
These three propagating degrees of freedom are not altered by the derivative interactions (\ref{L2})-(\ref{L6}), apart from the appearance of two tensor polarizations from 
the gravity sector \cite{Heisenberg:2014rta,Kimura:2016rzw}. 
The non-minimal derivative couplings (\ref{L4})-(\ref{L6}) are needed to keep the equations of motion up to second order. The gauge-invariant vector-tensor interaction introduced by 
Horndeski in 1976 corresponds to ${\cal L}=F+{\cal L}_4+{\cal L}_6$ with constant functions $G_4$ and $G_6$ \cite{Horndeski:1976gi}.

\subsection{Background equations of motion}

For the purpose of cosmological applications, we take
the flat FLRW metric with the line element 
$ds^2=-dt^2+a^2(t)\delta_{ij}dx^idx^j$, where $a(t)$ stands for the 
time-dependent scale factor with the cosmic time $t$. 
We consider the vector field $A^{\mu}$ with a 
time-dependent temporal component $\phi(t)$ alone, i.e., 
\begin{equation}
A^{\mu}=(\phi(t),0,0,0)\,,
\end{equation}
which is compatible with the background symmetry. 
Assuming that the matter field in the Lagrangian density 
${\cal L}_M$ (with energy density $\rho_M$ and pressure 
$P_M$) is minimally coupled to gravity, 
they obey the continuity equation 
\begin{equation}
\dot{\rho}_M+3H(\rho_M+P_M)=0\,,
\label{continuity}
\end{equation}
where $H \equiv \dot{a}/a$ is the Hubble expansion rate, 
and a dot represents a derivative with respect to $t$. 
Varying the action (\ref{LagProca}) with respect to
$g_{\mu \nu}$, we obtain the modified Einstein 
field equations 
\begin{eqnarray}
& &
G_2-G_{2,X}\phi^2-3G_{3,X}H \phi^3
+6G_4H^2-6(2G_{4,X}+G_{4,XX}\phi^2)H^2\phi^2+G_{5,XX} H^3\phi^5+ 5G_{5,X} H^3\phi^3
=\rho_M\,,
\label{be1}\\
& &
G_2-\dot{\phi}\phi^2G_{3,X}+2G_4\,(3H^2+2\dot{H})
-2G_{4,X}\phi \, ( 3H^2\phi +2H\dot{\phi}
+2\dot{H} \phi )\nonumber\\
&& {}-4G_{4,XX}H\dot{\phi}\phi^3+G_{5,XX}H^2\dot{\phi} \phi^4
+G_{5,X}
H \phi^2(2\dot{H}\phi +2H^2\phi+3H\dot{\phi})
=-P_M\,.
\label{be2}
\end{eqnarray}
Variation of the action (\ref{LagProca}) with respect to 
$\phi$ leads to 
\begin{equation}
\phi \left( G_{2,X}+3G_{3,X}H\phi +6G_{4,X}H^2
+6G_{4,XX}H^2\phi^2
-3G_{5,X}H^3\phi-G_{5,XX}H^3 \phi^3 \right)=0\,.
\label{be3}
\end{equation}
The functions $g_5,G_6$ and the additional dependence 
of $F$ and $Y$ in the function $G_2$, which correspond
to intrinsic vector modes, do not contribute to the background 
equations of motion as expected.
From Eq.~(\ref{be3}) one immediately observes that, 
for the branch $\phi \neq 0$,  
there exist interesting de Sitter solutions with 
constant values of $\phi$ and $H$ \cite{DeFelice:2016yws}.

\subsection{Concrete models}
\label{conmodelsec}

In a previous work, we have shown that viable dark energy models exist in the framework of generalized Proca theories \cite{DeFelice:2016yws}. 
As we mentioned before, the temporal vector component 
is not dynamical and can be expressed in terms of the Hubble 
parameter $H$. In order for the energy density of the temporal 
component $\phi$ to start dominating over the 
background matter densities at the late cosmological epoch, 
the amplitude of the field $\phi$ should increase with the
decrease of $H$. Thus, the relation should be of the form
\begin{equation}
\phi^p \propto H^{-1}\,,
\label{phiH}
\end{equation}
with a positive constant $p$. In the following, we assume that $\phi$ is positive. To guarantee the scaling (\ref{phiH}) between $\phi$ and $H$, 
the functions $G_{2,3,4,5}$ in Eq.~(\ref{be3}) 
should be chosen with the following specific scaling 
of $X$ \cite{DeFelice:2016yws}:
\begin{eqnarray}
& &
G_2(X)=b_2 X^{p_2}+F\,,\qquad
G_3(X)=b_3X^{p_3}\,,\qquad
G_4(X)=\frac{M_{\rm pl}^2}{2}+b_4X^{p_4}\,,\qquad
G_5(X)=b_5X^{p_5}\,,
\label{G2345}
\end{eqnarray}
with the powers $p_{3,4,5}$ of the form
\begin{equation}
p_3=\frac12 \left( p+2p_2-1 \right)\,,\qquad
p_4=p+p_2\,,\qquad
p_5=\frac12 \left( 3p+2p_2-1 \right)\,,
\label{p345}
\end{equation}
where $M_{\rm pl}$ is the reduced Planck mass and 
$b_{2,3,4,5}$ are constants.
Note that the specific case with $p_2=1$ and $p=1$ corresponds to 
vector Galileons \cite{Heisenberg:2014rta, Tasinato:2014eka},
where $\phi \propto H^{-1}$. 
The models given by the functions (\ref{G2345}) are the 
generalization of vector Galileons.

For the matter fields, we will assume the existence of 
non-relativistic matter (energy density $\rho_m$ and 
pressure $P_m=0$) and radiation (energy density $\rho_r$ and 
pressure $P_r=\rho_r/3$) together with their respective continuity equations
$\dot{\rho}_m+3H\rho_m=0$ and $\dot{\rho}_r+4H\rho_r=0$.
Then we have $\rho_M=\rho_m+\rho_r$ and $P_M=\rho_r/3$ 
in Eqs.~(\ref{be1}) and (\ref{be2}), respectively. 
For later convenience, we define the following 
dimensionless quantities (where $i=3,4,5$):
\begin{equation}
y \equiv \frac{b_2\phi^{2p_2}}
{3M_{\rm pl}^2 H^2\,2^{p_2}}\,,\qquad
\beta_i \equiv
\frac{p_ib_i}{2^{p_i-p_2}p_2b_2}
\left( \phi^p H \right)^{i-2}\,,
\label{ydef}
\end{equation}
and the density parameters
\begin{equation}
\Omega_{r} \equiv
\frac{\rho_r}{3M_{\rm pl}^2H^2}\,,
\qquad
\Omega_{m} \equiv
\frac{\rho_m}{3M_{\rm pl}^2H^2}\,,
\qquad
\Omega_{\rm DE} \equiv
1-\Omega_{r}-\Omega_{m}\,.
\end{equation}
The variables $\beta_i$'s are constants due to the relation (\ref{phiH}). 
For the branch $\phi \neq 0$, Eq.~(\ref{be3}) is expressed in a simple form 
\begin{equation}
1+3\beta_3+6(2p+2p_2-1)\beta_4
-(3p+2p_2)\beta_5=0\,,
\label{mueq}
\end{equation}
which can be exploited to express $\beta_3$ in terms of $\beta_4$ 
and $\beta_5$. On using Eq.~(\ref{be1}), the dark energy density 
parameter is related to the quantity $y$, as
\begin{equation}
\Omega_{\rm DE}=
\frac{\beta y}{p_2(p+p_2)}\,,
\label{OmegaDE}
\end{equation}
where the constant $\beta$ is defined by 
\begin{equation}
\beta \equiv -p_2(p+p_2)(1+4p_2 \beta_5)
+6p_2^2(2p+2p_2-1)\beta_4\,.
\label{betadef}
\end{equation}
For the constant $b_2$ appearing in $G_2(X)$, we choose it to be negative, i.e., 
$b_2=-m^2M_{\rm pl}^{2(1-p_2)}$, where $m$ is a mass term. 
This is for avoiding the appearance of tensor ghosts in the limit 
that $G_5 \to 0$ \cite{DeFelice:2016yws}. We also introduce the 
following dimensionless quantity 
\be
\lambda \equiv \left( \frac{\phi}{M_{\rm pl}} \right)^p 
\frac{H}{m}\,,
\ee
which is constant from Eq.~(\ref{phiH}). On using 
Eqs.~(\ref{ydef}) and (\ref{betadef}), the temporal vector 
component can be expressed as 
$\phi=M_{\rm pl}[-2^{p_2}\cdot 3\lambda^2 p_2(p+p_2)
\Omega_{\rm DE}/\beta]^{1/[2(p+p_2)]}$. 
We will focus on the case $\beta<0$, under which $\phi>0$ 
and $\lambda>0$ for $p_2(p+p_2)>0$.

To bring the equations of motion into an autonomous form, we perform 
the following manipulations: first we differentiate the $\phi \neq 0$ branch 
of Eq.~(\ref{be3}) with respect to $t$ and together with Eq.~(\ref{be2}) 
we can solve them for $\dot{\phi}$ and $\dot{H}$.
Taking the derivatives of $\Omega_{\rm DE}$ and $\Omega_{r}$ 
with respect to ${\cal N} \equiv \ln a$ (denoted by a prime), 
we obtain the following autonomous equations
\begin{eqnarray}
\Omega_{\rm DE}' &=& \frac{(1+s)\Omega_{\rm DE}
(3+\Omega_r-3\Omega_{\rm DE})}
{1+s\,\Omega_{\rm DE}}\,,
\label{dx1} \\
\Omega_{r}' &=& -\frac{\Omega_r [ 1-\Omega_r
+(3+4s)\Omega_{\rm DE}]}
{1+s\,\Omega_{\rm DE}}\,,
\label{dx2}
\end{eqnarray}
where we introduced the variable 
\begin{equation}
s \equiv \frac{p_2}{p}\,.
\end{equation}
After integrating Eqs.~(\ref{dx1}) and (\ref{dx2}) for given initial conditions of $\Omega_{\rm DE}$ and $\Omega_r$, the three density 
parameters $\Omega_{\rm DE}, \Omega_r$ and 
$\Omega_m=1-\Omega_{\rm DE}-\Omega_{r}$ are known 
accordingly. Furthermore, we impose the condition $s>-1$ 
to prevent $\Omega_{\rm DE}$ from diverging in the interval
$0\leq\Omega_{\rm DE}\leq 1$. We also define the effective equation of state of the system by $w_{\rm eff} \equiv -1-2\dot{H}/(3H^2)$,
which can be expressed as
\begin{equation}
w_{\rm eff}=
\frac{\Omega_r-3(1+s)\Omega_{\rm DE}}
{3(1+s\,\Omega_{\rm DE})}\,.
\label{weff}
\end{equation}

We write Eqs.~(\ref{be1}) and (\ref{be2}) in the forms 
$3M_{\rm pl}^2H^2=\rho_{\rm DE}+\rho_M$ and 
$M_{\rm pl}^2 (3H^2+2\dot{H})=-P_{\rm DE}-P_M$, 
respectively, where
\begin{eqnarray}
\rho_{\rm DE} &=&
-G_2+G_{2,X}\phi^2+3G_{3,X}\phi^3 H
-6g_4H^2+6\phi^2H^2 (2G_{4,X}+G_{4,XX}\phi^2)-H^3 G_{5,XX}\phi^5-5H^3G_{5,X} \phi^3,
\label{rhode}\\
P_{\rm DE} &=&
G_2-\dot{\phi}\phi^2G_{3,X}+2g_4\,(3H^2+2\dot{H})
-2\phi G_{4,X}\, ( 3\phi H^2 +2\dot{\phi}H
+2\phi \dot{H} )\nonumber\\
&& {}-4H\dot{\phi}\phi^3G_{4,XX}+\dot\phi\phi^4H^2G_{5,XX}+G_{5,X}\phi^2H(2\phi\dot H+2\phi H^2+3\dot\phi H)\,,
\label{Pde}
\end{eqnarray}
with $g_4(X)=b_4X^{p_4}$.
Defining the dark energy equation of state as
$w_{\rm DE}=P_{\rm DE}/\rho_{\rm DE}$, 
it follows that 
\begin{equation}
w_{\rm DE}=
-\frac{3(1+s)+s\,\Omega_r}
{3(1+s\,\Omega_{\rm DE})}\,.
\label{wde}
\end{equation}
Using Eqs.~(\ref{dx1}) and (\ref{dx2}), we obtain 
a single differential equation of the form
\begin{equation}
\frac{\Omega_{\rm DE}'}{\Omega_{\rm DE}}
=(1+s) \left( \frac{\Omega_{r}'}{\Omega_{r}}+4 \right)\,.
\end{equation}
This equation is easily integrated to give 
\begin{equation}
\frac{\Omega_{\rm DE}}{\Omega_{r}^{1+s}}
=\frac{\Omega_{\rm DE0}}{\Omega_{r0}^{1+s}}
\left( \frac{a}{a_0} \right)^{4(1+s)}\,,
\label{Omere}
\end{equation}
where the lower subscript ``0'' denotes today's values. 

\section{Cosmological perturbations}
\label{persec}

By considering linear cosmological perturbations on the flat 
FLRW background, the conditions for avoiding ghosts and Laplacian 
instabilities in the small-scale limit were already derived 
in Refs.~\cite{DeFelice:2016yws,DeFelice:2016uil}. 
Here, we briefly review six no-ghost and stability conditions 
arising from tensor, vector, and scalar perturbations. 
We also discuss the equations of motion for matter perturbations 
to confront generalized Proca theories with the observations 
of RSD in Sec.~\ref{perconsec}.

\subsection{Stability conditions}

For the metric we take the perturbed line element in 
the flat gauge
\be
ds^{2}=-(1+2\alpha)\,dt^{2}+2\left( \partial_{i}\chi
+V_i \right)dt\,dx^{i}+a^{2}(t) \left( \delta_{ij}
+h_{ij} \right)dx^i dx^j\,,
\ee
where $\alpha, \chi$ are scalar metric perturbations,
$V_i$ is the vector perturbation obeying the transverse
condition $\partial^i V_i=0$, and $h_{ij}$ is
the tensor perturbation satisfying the transverse and
traceless conditions $\partial^i h_{ij}=0$ and
${h_i}^i=0$. The temporal and spatial components 
of the vector field can be decomposed into the 
background and perturbed parts, as 
\be
A^{0}=\phi(t)+\delta\phi\,,\qquad
A^{i}=\frac{1}{a^2(t)} \delta^{ij} \left(
\partial_{j}\chi_{V}+E_j \right)\,,
\ee
where $\delta \phi$ and $\chi_{V}$ are scalar perturbations, 
and $E_j$ is the vector perturbation satisfying 
$\partial^j E_j=0$. 
The Schutz-Sorkin action \cite{Sorkin} allows us to describe 
both vector and scalar perturbations of the matter perfect fluid.
For scalar perturbations, the key observables are the matter 
density perturbation $\delta \rho_M$ and the velocity potential $v$.

First of all, the tensor perturbation $h_{ij}$ has two polarization modes 
$h_{+}$ and $h_{\times}$, which can be expressed as 
$h_{ij}=h_{+}e_{ij}^{+}+h_{\times}e_{ij}^{\times}$
in terms of the unit tensors obeying 
the normalizations $e_{ij}^{+}({\bm k}) e_{ij}^{+}(-{\bm k})^*=1$,
$e_{ij}^{\times}({\bm k}) e_{ij}^{\times}(-{\bm k})^*=1$,
$e_{ij}^{+}({\bm k}) e_{ij}^{\times}(-{\bm k})^*=0$.
Expanding the action (\ref{LagProca}) in $h_{ij}$ up
to quadratic order, the second-order action yields 
\be
S_T=\sum_{\lambda={+},{\times}}\int dt\,d^3x\,
a^3\,\frac{q_T}{8}  \left[\dot{h}_\lambda^2
-\frac{c_T^2}{a^2}(\partial h_\lambda)^2\right]\,.
\label{ST}
\ee
The quantities $q_T$ and $c_T^2$ determine no-ghost and 
stability conditions, respectively, whose explicit forms 
are given by 
\ba
q_T 
&=& 2G_4-2\phi^{2}G_{{4,X}}+
H\phi^{3}G_{{5,X}}>0\,, 
\label{qT} \\
c_{T}^{2}
&=& \frac {2G_{4}+\phi^{2}\dot\phi\,G_{{5,X}}}
{q_T}>0\,.
\label{cT}
\ea

For vector perturbations, the dynamical field is given by  
the combination 
\be
Z_i=E_i+\phi(t)V_i\,.
\ee
Due to the transverse condition $\partial^i Z_i=0$, 
there are two propagating degrees of freedom for $Z_i$, 
e.g., $Z_i=(Z_1(z), Z_2(z), 0)$ for the vector field whose 
wavenumber ${\bm k}$ is along the $z$ direction. 
Expanding the action (\ref{LagProca}) up to quadratic order 
and taking the small-scale limit, the resulting second-order 
action for $Z_1$ and $Z_2$ can be written as the form 
analogous to Eq.~(\ref{ST}) with the no-ghost 
and stability conditions
\ba
\hspace{-0.7cm}
q_V 
&=& G_{2,F}+2G_{2,Y}\phi^2-4g_5H \phi
+2G_6H^2+2G_{6,X} H^2 \phi^2>0\,, 
\label{qV} \\
\hspace{-0.7cm}
c_{V}^{2}
&=& 1+\frac{\phi^2(2G_{4,X}-G_{5,X}H \phi)^2}
{2q_Tq_V}+\frac{2[G_6 \dot{H}-G_{2,Y}\phi^2
-(H\phi-\dot{\phi})(H \phi\,G_{6,X}-g_5)]}
{q_V}>0\,.
\label{cV}
\ea

For scalar perturbations, the dynamical field arising from 
the vector field is given by 
\be
\psi=\chi_V+\phi(t) \chi\,.
\ee
The matter perturbation $\delta \rho_M$, which arises from 
the Schutz-Sorkin action, also works as a dynamical scalar field.
The second-order action of scalar perturbations is given in 
Eq.~(4.6) of Ref.~\cite{DeFelice:2016uil}.
Varying this action with respect to 
$\alpha, \chi, \delta \phi, \partial \psi,v,$ and $\delta \rho_M$, 
the equations of motion in Fourier space 
are given, respectively, by 
\ba
& &
\delta \rho_M-2w_4 \alpha+\left( 3Hw_1-2w_4 \right)\frac{\delta \phi}{\phi}
+\frac{k^2}{a^2} \left( {\cal Y}
+w_1 \chi-w_6 \psi \right)=0\,,\label{per1} \\
& &
\left( \rho_M+P_M \right) v+
w_1 \alpha+\frac{w_2}{\phi} \delta \phi=0\,,\label{per2}\\
& &
\left( 3Hw_1-2w_4 \right)\alpha-2w_5 \frac{\delta \phi}{\phi}
+\frac{k^2}{a^2} \left[ \frac12 {\cal Y}
+w_2 \chi-\frac12 \left( \frac{w_2}{\phi}+w_6 \right) \psi
\right]=0\,,\label{per3} \\
& &
\dot{\cal Y}+\left( H -\frac{\dot{\phi}}{\phi} \right){\cal Y}
+2\phi \left( w_6 \alpha+w_7 \psi \right)
+\left( \frac{w_2}{\phi}+w_6 \right) \delta \phi
=0\,,\label{per4}\\
& &
\dot{\delta \rho}_M+3H\left(1+c_M^2 \right) \delta \rho_M
+\frac{k^2}{a^2} \left( \rho_M+P_M \right)
\left( \chi+v \right)=0\,,\label{per5}\\
& &
\dot{v}-3Hc_M^2 v-c_M^2 \frac{\delta \rho_M}
{\rho_M+P_M}-\alpha=0\,,\label{per6}
\ea
where $c_M^2$ is the matter propagation speed squared, 
and
\ba
w_{1} & = & {H}^{2} {\phi}^{3} (G_{{5,X}}+{\phi}^{2}G_{{5,{\it XX}}})
-4\,H(G_{{4}}+{\phi}^{4}G_{{4,{\it XX}}})-{\phi}^{3}G_{{3,X}}\,,
\label{w1}\\
w_{2} & = & w_1+2Hq_T\,,\label{w2}\\
w_{3} & = & -2{\phi}^{2}q_V\,,\label{w3} \\
w_{4} & = & \frac{1}{2}{H}^{3}\phi^{3}(9G_{{5,X}}-\phi^{4}G_{{5,{\it XXX}}})
-3\,H^{2} (2G_{{4}}+2\phi^{2}G_{{4,X}}+\phi^{4}G_{{4,{\it XX}}}-\phi^{6}G_{{4,{\it XXX}}}) \nonumber \\
 & &-\frac{3}{2}\,H\phi^{3}(G_{{3,X}}-\phi^{2}G_{{3,{\it XX}}})
 +\frac{1}{2}\,\phi^{4}G_{{2,{\it XX}}}\,,\label{w4} \\
w_{5} & = & w_{4}-\frac{3}{2}\,H(w_{1}+w_{2})\,, \label{w5} \\
w_{6} & = & -\phi \left[{H}^{2}\phi(G_{{5,X}}-{\phi}^{2}G_{{5,{\it XX}}})
-4\,H(G_{{4,X}}-{\phi}^{2}G_{{4,{\it XX}}})+\phi G_{{3,X}}\right]\,, \label{w6} \\
w_{7} & = & 2(H\phi G_{{5,X}}-2G_{{4,X}}) \dot{H}
+\left[H^{2}(G_{{5,X}}+{\phi}^{2}G_{{5,{\it XX}}})-4\,H\phi\,G_{{4,{\it XX}}}-G_{{3,X}}\right] \dot{\phi}\,,
\label{w7}\\
{\cal Y} &=& \frac{w_3}{\phi}
\left( \dot{\psi}+\delta \phi+2\alpha \phi \right)\,.
\label{Ydefi}
\ea
On using Eqs.~(\ref{per1})-(\ref{per3}) and (\ref{per5}), 
we can express $\alpha,\chi,\delta \phi,v$ in terms of 
$\psi, \delta \rho_M$ and their derivatives. 
Then, the second-order action of scalar perturbations is
written in terms of the two dynamical fields 
$\psi$ and $\delta \rho_M$. 
This allows one to derive no-ghost and stability conditions 
in the small-scale limit. 
The no-ghost and stability conditions of the matter 
field $\delta \rho_M$ are trivially satisfied for 
$\rho_M+P_M>0$ and $c_M^2>0$.
For the perturbation $\psi$, we require the 
following conditions \cite{DeFelice:2016yws,DeFelice:2016uil}
\ba
Q_{S}
&=& \frac{a^{3}H^2q_Tq_S}
{\phi^{2}(w_{1}-2w_{2})^{2}}>0\,,
\label{Qsge}\\
c_{S}^2
&=&
\frac{\mu_S}{8H^2 \phi^2 q_Tq_Vq_S}>0\,,
\label{csge}
\ea
where 
\ba
q_S &=& 3w_1^2+4q_Tw_4\,,\\
\mu_S &=& \left[ w_6 \phi (w_1-2w_2)
+w_1 w_2 \right]^2-w_3 \left( 2w_2^2 \dot{w}_1
-w_1^2 \dot{w}_2 \right)+\phi \left( w_1-2w_2 \right)^2 w_3 \dot{w}_6 
\nonumber \\
& & +w_3(w_1-2w_2) \left[ \left( H -2\dot{\phi}/\phi \right)
w_1 w_2+\left( w_1-2w_2 \right)
\left\{ w_6 \left( H \phi -\dot{\phi} \right)+2w_7 \phi^2
\right\} \right] \nonumber \\
& &
+2w_2^2 w_3 \left( \rho_M+P_M \right)\,.
\label{muc}
\ea
Under the no-ghost conditions of tensor and vector perturbations 
($q_T>0, q_V>0$), Eqs.~(\ref{Qsge}) 
and (\ref{csge}) can be satisfied for $q_S>0$ and 
$\mu_S>0$. There are specific cases where the quantity 
$w_1-2w_2$ in Eq.~(\ref{Qsge}) crosses 0, at which $Q_S$ exhibits 
the divergence \cite{DeFelice:2016yws}. 
We will exclude such cases for constraining the 
viable parameter space.

\subsection{Effective gravitational coupling with 
matter perturbations}

To confront generalized Proca theories with the observations 
of large-scale structures and weak lensing, we consider 
non-relativistic matter (labeled by $m$) with the equation of state 
$w_m=P_m/\rho_m=0^{+}$ 
and the sound speed squared $c_m^2=0^{+}$.
We introduce the matter density contrast $\delta$
and the gauge-invariant gravitational potentials 
\be
\delta=\frac{\delta \rho_m}{\rho_m}+3Hv\,,\qquad
\Psi=\alpha+\dot{\chi}\,,\qquad 
\Phi=H\chi\,.
\ee
Taking the time derivative of Eq.~(\ref{per5}) and 
using Eq.~(\ref{per6}), we obtain 
\be
\ddot{\delta}+2H\dot{\delta}+\frac{k^2}{a^2}\Psi
=3\ddot{\cal B}+6H \dot{\cal B}\,,
\label{deleq}
\ee
where ${\cal B} \equiv Hv$. 
The effective gravitational coupling $G_{\rm eff}$ 
is defined by 
\be
\frac{k^2}{a^2}\Psi=-4\pi G_{\rm eff} \rho_m \delta\,.
\label{Geff}
\ee

For the perturbations deep inside the sound horizon 
($c_S^2k^2/a^2 \gg H^2$), we can resort to the quasi-static 
approximation for deriving the relations between 
$\Psi,\Phi$ and $\delta$ \cite{Boi,Tsuji07,DKT}.
Provided that $c_S^2$ is not very much smaller than 1,
the dominant contributions to the perturbation equations 
of motion are the terms containing $k^2/a^2$ and 
$\delta \rho_m$. The terms on the r.h.s. of Eq.~(\ref{deleq}) 
are negligible relative to those on the l.h.s., so that 
\be
\ddot{\delta}+2H\dot{\delta}
-4\pi G_{\rm eff} \rho_m \delta
\simeq 0\,.
\label{deleq2}
\ee
On using the quasi-static approximation for 
Eqs.~(\ref{per1})-(\ref{per4}), the effective 
gravitational coupling is analytically known as 
\cite{DeFelice:2016uil}
\be
G_{\rm eff}
= \frac{\xi_2+\xi_3}{\xi_1}\,,
\label{Geff2}
\ee
where 
\ba
\hspace{-0.8cm}
\xi_1 &=& 4\pi \phi^2 \left( w_2+2H q_T \right)^2\,,
\label{xi1} \\
\hspace{-0.8cm}
\xi_2 &=& \left[ H\left( w_2+2Hq_T \right)-\dot{w}_1
+2\dot{w}_2+\rho_m \right]\phi^2-\frac{w_2^2}
{q_V}\,,\label{xi2}\\
\hspace{-0.8cm}
\xi_3 &=& \frac{1}{8H^2 \phi^2 q_S^3 q_T c_S^2}
\biggl[ 2\phi^2 \left\{ q_S [w_2\dot{w}_1
-(w_2-2Hq_T) \dot{w}_2]+\rho_m w_2
[3w_2(w_2+2Hq_T)-q_S] \right\} \nonumber \\
& &~~~~~~~~~~~~~~~~~~~
+\frac{q_S}{q_V}w_2
\left\{ w_2 (w_2-2Hq_T)-w_6 \phi (w_2+2Hq_T) \right\}
\biggr]^2\,.
\label{xi3} 
\ea
The solutions to $\delta$ derived by solving Eq.~(\ref{deleq2}) with Eq.~(\ref{Geff2}) can reproduce full numerical results 
at high accuracy \cite{DeFelice:2016uil}.

Besides $G_{\rm eff}$, the gravitational slip parameter 
$\eta=-\Phi/\Psi$ is also an important quantity for describing 
the deviation of light rays in weak lensing 
observations \cite{Sapone}.  
Under the quasi-static approximation 
it follows that \cite{DeFelice:2016uil}
\be
\eta=\frac{\xi_4}{\xi_2+\xi_3}\,,\label{eta2}
\ee
where 
\ba
\xi_4 &=& \frac{w_2+2Hq_T}{4H q_S^2q_Vq_Tc_S^2}
\biggl[ 4H^2 \phi^2 q_S^2q_Vq_Tc_S^2+
2\phi^2 q_S q_V w_2\dot{w}_2 (w_2-2Hq_T)
+w_2^2\{\phi q_S w_6(w_2+2Hq_T) \nonumber \\
& &
-w_2 q_S(w_2-2Hq_T)
-2\phi^2 q_S q_V \dot{w}_1
+2\phi^2 q_V [q_S-3w_2(w_2+2Hq_T)] \rho_m \} \biggr]\,.
\label{xi4}
\ea
Although we do not use the information of the gravitational slip parameter $\eta$ in 
our likelihood analysis, this can be important for 
confronting the model with
future high-precision observations of weak lensing.

\section{Observational data}
\label{datasec}

We would like to test for the model (\ref{G2345}) with different 
observational data from CMB, BAO, SN Ia, the Hubble expansion rate, 
and RSD measurements. 
In this section, we will explain the data sets used in the likelihood 
analysis performed in Secs.~\ref{backconsec} and \ref{perconsec}.

\subsection{CMB}

The CMB power spectrum is affected by the presence of dark energy in at least two ways \cite{Copeland:2006wr}. 
First, the positions of CMB acoustic peaks are shifted 
by the change of the angular diameter distance from the last 
scattering surface to today. 
Second, the variation of gravitational potentials induced by the presence 
of dark energy leads to the late-time integrated Sachs-Wolfe effect. 
The latter mostly affects the temperature anisotropies on large scales, at which the observational data are prone to uncertainties induced 
by the cosmic variance.
Since the first effect is usually more important to constrain the property of 
dark energy, we will focus on the CMB distance measurements in the following.

The comoving distance to the CMB decoupling surface $r(z_*)$ 
(the redshift $z_* \simeq 1090$) and the comoving sound horizon 
at the CMB decoupling $r_{\rm s}(z_*)$ can be constrained from 
CMB measurements. In particular, the CMB shift parameters 
\ba
{\cal R} &=& \sqrt{\Omega_{m0}}\,H_0 r(z_*)\,,
\label{shift1}\\
l_a &=& \frac{\pi r(z_*)}{r_{\rm s}(z_*)}
\label{shift2}
\ea
are the two key quantities for placing constraints 
on dark energy \cite{Wang1,Komatsu:2008hk,Wang2}. 
The shift parameter ${\cal R}$ is associated with the overall amplitude of 
CMB acoustic peaks, whereas $l_a$ determines the average acoustic 
structure. We need to employ both ${\cal R}$ and $l_a$ for extracting 
the necessary information to constrain dark energy models 
from the CMB power spectrum.

The comoving distance on the flat FLRW background is given by 
$r(z_*)=\int_0^{z_*}dz/H(z)$, where $z=a_0/a-1$ is the redshift.
Then, Eq.~(\ref{shift1}) reads
\begin{equation}
\mathcal{R}=\sqrt{\Omega_{m0}}\int_{0}^{z_{*}}\frac{dz}{E(z)}\,,
\end{equation}
where $E(z)$ represents the Hubble ratio given by
\begin{equation}
E(z) \equiv \frac{H(z)}{H_{0}}
=\sqrt{\frac{\Omega_{r0}}{\Omega_{r}}}(1+z)^{2}\,.
\label{Ez}
\end{equation}
We can promote $\mathcal{R}$ to a function of $z$ 
satisfying the condition $\mathcal{R}(0)=0$.
Defining the ratios $\bar{\Omega}_{r}=\Omega_{r}/\Omega_{r0}$ 
and $\bar{z}=z/z_*$, it follows that 
\begin{equation}
\frac{d\mathcal{R}(\bar{z})}{d\bar{z}}  = \frac{z_{*}\sqrt{\Omega_{m0}\bar{\Omega}_{r}}}{(1+z_{*}\bar{z})^{2}}\,,
\label{eqRandR0}
\end{equation}
which should be integrated from 
$\bar{z}=0$ to $\bar{z}=1$ with $\mathcal{R}(0)=0$.
In doing so, we need to know $\bar{\Omega}_{r}$ as a function 
of $\bar{z}$. For the model (\ref{G2345}) with Eq.~(\ref{p345}), 
the dark energy density parameter is known from 
Eq.~(\ref{Omere}), as
\begin{equation}
\Omega_{\rm DE}=\left( 1-\Omega_{m0}-\Omega_{r0} \right) 
\bar{\Omega}_{r}^{1+s}(1+z_{*}\bar{z})^{-4(1+s)}\,.
\label{OmeDE2}
\end{equation}
On using Eq.~(\ref{dx2}), the radiation density parameter obeys
\begin{equation}
\frac{d\bar{\Omega}_r}{d\bar{z}}
=\frac{z_* \bar{\Omega}_r[1-\Omega_{r0}\bar{\Omega}_r
+(3+4s)\left( 1-\Omega_{m0}-\Omega_{r0} \right) 
\bar{\Omega}_{r}^{1+s}(1+z_{*}\bar{z})^{-4(1+s)}]}
{(1+z_*\bar{z})[1+s\left( 1-\Omega_{m0}-\Omega_{r0} \right) 
\bar{\Omega}_{r}^{1+s}(1+z_{*}\bar{z})^{-4(1+s)}]}\,,
\label{Omereq2}
\end{equation}
with $\bar{\Omega}_r(0)=1$.
We integrate Eqs.~(\ref{eqRandR0}) and (\ref{Omereq2}) to 
compute the value of ${\cal R}$ at $\bar{z}=1$.

The dimensionless comoving distance 
$\bar{r}(z)=H_0 r(z)$, where $r(z)=\int_0^z dz'/H(z')$, 
obeys the differential equation 
\begin{equation}
\frac{d\bar{r}(\bar{z})}{d\bar{z}}=\frac{z_* \sqrt{\bar{\Omega}_r}}
{(1+z_{*}\bar{z})^2}\,,
\label{codiff}
\end{equation}
where $\bar{r}(0)=0$.
The comoving sound horizon at the redshift $z$ is defined by 
\begin{equation}
r_{\rm s}(z)=\int_0^{t} \frac{c_{\rm s}\,dt}{a}=
\frac{1}{H_{0}}\,\int_{z}^{\infty}dz' 
\frac{c_{\rm s}(z')}
{E(z')}\,,
\label{rsz}
\end{equation}
where we used the normalization $a_0=1$ in the second equality.
The sound speed squared $c_{\rm s}$ of the coupled system of 
baryons (density $\rho_b$) and photons (density $\rho_{\gamma}$) 
is given by \cite{Sugiyama}\footnote{The quantity $c_{\rm s}$ is different from 
the sound speed $c_S$ of the scalar degree of freedom $\psi$ arising 
from the vector field. Since the density of the vector field is much smaller 
than those of the background fluids before the CMB decoupling epoch, 
the existence of the vector field does not affect $c_{\rm s}$.}
\be
c_{\rm s}=\frac{1}{\sqrt{3[1+R_{b}/(1+z)]}}\,,
\label{cs2}
\ee
where 
\be
R_b=\frac{3\rho_{b0}}{4\rho_{\gamma 0}}
=31500\Omega_{b0}\,h^{2}\left(\frac{2.7255}{2.7}\right)^{-4}\,.
\ee
Introducing the dimensionless quantity $\bar{r}_{\rm s}=H_{0}r_{\rm s}$ 
and taking the $z$ derivative of Eq.~(\ref{rsz}), we have  
$d\bar{r}_{\rm s}/dz=-c_{\rm s}(z)/E(z)$. 
Upon the change of variable, $\bar{a}=a/a_*$, it follows that 
$1+z=1/a=1/(a_{*}\bar{a})=(1+z_{*})/\bar{a}$.
On using Eqs.~(\ref{Ez}) and (\ref{cs2}), the dimensionless 
distance $\bar{r}_{\rm s}$ obeys the differential equation 
\begin{equation}
\frac{d\bar{r}_{\rm s}}{d\bar{a}} 
=\frac{1}{1+z_{*}}\,\frac{\sqrt{\Omega_{r}(\bar{a})}}{\sqrt{3\Omega_{r0}[1+R_{b}\bar{a}/(1+z_{*})]}}\,,
\label{drseq}
\end{equation}
with $r_{\rm s}(\bar{a}=0)=0$.

The radiation density parameter is known from Eq.~(\ref{dx2}), such that 
\begin{equation}
\frac{d\Omega_r}{d\bar{a}} 
=-\frac{\Omega_r [ 1-\Omega_r
+(3+4s)\Omega_{\rm DE}]}
{\bar{a}(1+s\,\Omega_{\rm DE})}\,.
\label{dOmereq}
\end{equation}
At first glance the r.h.s. of Eq.~(\ref{dOmereq}) looks divergent 
in the limit $\bar{a} \to 0$, but this is not the case because the 
numerator also approaches 0. 
For numerical purposes, we will rewrite the r.h.s. of Eq.~(\ref{dOmereq}) 
in a more convenient form. In doing so, we introduce the following quantity 
\begin{equation}
\Gamma \equiv
\frac{\rho_{r}}{\rho_{r}+\rho_{m}}=\frac{\Omega_{r}}{\Omega_{r}+\Omega_{m}}=\frac{\Omega_{r}}{1-\Omega_{{\rm DE}}}
=\frac{\Omega_{r0}(1+z_*)}
{\Omega_{r0}(1+z_*)+\Omega_{m0}\bar{a}}\,.
\end{equation}
Since $\Omega_{r}=\Gamma(1-\Omega_{{\rm DE}})$, the quantity 
$1-\Omega_r$ in Eq.~(\ref{dOmereq}) can be expressed as
\begin{equation}
1-\Omega_{r}=
1-\Gamma+\Gamma\Omega_{{\rm DE}}
= \frac{\Omega_{m0}\bar{a}}{\Omega_{r0}(1+z_*)
+\Omega_{m0}\bar{a}}+\frac{\Omega_{r0}(1+z_*)}
{\Omega_{r0}(1+z_*)+\Omega_{m0}\bar{a}}\Omega_{\rm DE}\,,
\end{equation}
where $\Omega_{\rm DE}=(1-\Omega_{m0}-\Omega_{r0})(1+z_*)^{-4(1+s)}
(\Omega_r/\Omega_{r0})^{1+s}\bar{a}^{4(1+s)}$ from Eq.~(\ref{Omere}).
Then, we can express Eq.~(\ref{dOmereq}) in the form 
\begin{eqnarray}
\frac{d\Omega_{r}}{d\bar{a}}
&=& -\frac{\Omega_{r}}{[1+s(1-\Omega_{m0}-\Omega_{r0})(\Omega_{r}/\Omega_{r0})^{1+s}(1+z_{*})^{-4(1+s)}
\bar{a}^{4(1+s)}]}\nonumber \\
& & {}\times\left\{\frac{\Omega_{m0}}{\Omega_{r0}(1+z_{*})+\Omega_{m0}\bar{a}}+\left[\frac{\Omega_{r0}(1+z_{*})}{\Omega_{r0}(1+z_{*})+\Omega_{m0}\bar{a}}+3+4s\right] 
\frac{(1-\Omega_{m0}-\Omega_{r0})\Omega_{r}^{1+s}\bar{a}^{3+4s}}{(1+z_{*})^{4(1+s)}\Omega_{r0}^{1+s}}\right\}, 
\label{dOmereq2} 
\label{Omera}
\end{eqnarray}
with $\Omega_{r}(\bar{a} \to 0) = 1$.
Provided that $3+4s>0$ we have $d\Omega_{r}/d\bar{a}\to-\Omega_{m0}/[\Omega_{r0}(1+z_{*})]$ as 
$\bar{a} \to 0$, so the r.h.s. of Eq.~(\ref{dOmereq2}) 
remains finite. 
Integrating Eqs.~(\ref{drseq}) and (\ref{Omera}) with Eq.~(\ref{codiff}), 
the second CMB shift parameter $l_a=\pi \bar{r}(z_*)/\bar{r}_{\rm s}(z_*)$ 
can be computed accordingly.

The CMB shift parameters extracted from the 
Planck 2015 data have the 
mean values $\langle l_a \rangle=301.77$ and 
$\langle {\cal R} \rangle=1.7482$ with the deviations 
$\sigma(l_a)=0.090$ and $\sigma({\cal R})=0.0048$,
respectively \cite{Wang3}. 
We fix the baryon density parameter as $\Omega_{b} h^2=0.02226$, 
where $h$ is the normalized Hubble constant
($H_0=100\,h$ km\,s$^{-1}$\,Mpc$^{-1}$).
The components of the normalized covariance matrix ${\bm C}$
are given by $C_{11}=C_{22}=1$ and $C_{12}=C_{21}=0.3996$. 
Then, the $\chi^{2}$ statistics associated with the CMB 
shift parameters is given by\footnote{The Planck 2015 team \cite{Ade:2015rim} 
provided the values $l_a =301.76 \pm 0.14$
and ${\cal R} =1.7488 \pm 0.0074$ 
with the covariance matrix components 
$C_{11}=C_{22}=1$ and $C_{12}=C_{21}=0.54$. 
We confirm that the corresponding $\chi_{\rm CMB}^{2}$ analysis gives very similar likelihood results as those 
derived from Eq.~(\ref{chicmb}).
} 
\begin{equation}
\chi_{\rm CMB}^{2} 
=(l_{a}-301.77)^{2}\times146.916+
({\cal R}-1.7482)^{2}\times51650.31
+2(\mathcal R-1.7482)(l_{a}-301.77)\times(-1100.77)\,.
\label{chicmb}
\end{equation}
The dependence on the parameters $\Omega_{m0}$, $h$ and $s$ are encoded in 
${\cal R}$ and $l_{a}$. For the test parameters 
$\Omega_{m0}=0.3$, $h=0.6774$, and $s=0.1$, 
we find that $\chi_{\rm CMB}^{2}=1519.87$.

\subsection{BAO}
\label{sec:BAO}

We also use the BAO data to constrain our model further. 
The BAO represent periodic fluctuations of the density of baryonic matter as a result of the counteracting forces of pressure and gravity. The photons release a pressure after the decoupling, which on the other hand creates a shell of baryonic matter at the sound horizon. {}From the BAO measurements we can deduce the distance-redshift relation at the observed redshifts. One of the important quantities is the sound horizon $r_{\rm s}(z_d)$, where $z_d$ is the redshift at which baryons are released from photons.
There is a fitting formula for the drag redshift $z_d$, 
as \cite{Eisenstein:1997ik}
\begin{equation}
z_d=\frac{1291(\Omega_{m0}h^2)^{0.251}}{1+0.659(\Omega_{m0}h^2)^{0.828}} \left[ 1+b_1(\Omega_bh^2)^{b_2} \right] \,,
\end{equation}
where the parameters $b_1$ and $b_2$ are
\begin{eqnarray}
b_1&=&0.313(\Omega_{m0}h^2)^{-0.419}
\left[ 1+0.607(\Omega_{m0}h^2)^{0.674} \right]\,,  \\
b_2&=&0.238 (\Omega_{m0}h^2)^{0.223} \,.
\end{eqnarray}

The sound horizon $r_{\rm s}(z)$ is given by Eq.~(\ref{rsz}) 
with the sound speed (\ref{cs2}). 
We perform a change of the variable $\tilde{a}=a/a_d$ with 
$1+z=a_0/(a_d\tilde{a})=(1+z_d)/\tilde{a}$. 
This helps us to make use of some of the results in the CMB 
distance measurements. 
Since $\tilde{a}/\bar{a}=(1+z_d)/(1+z_{*})$, it follows 
that $\bar{a}(a=a_d)=(1+z_*)/(1+z_d)$. 
With this change of variables, we can now integrate 
Eq.~(\ref{dOmereq2}) from $\bar{a}=0$ to $\bar{a}=(1+z_{*})/(1+z_d)$.
For the test parameters $\Omega_{m0}=0.3$, $h=0.6774$ and $s=0.1$, 
we obtain the value $\bar{r}_{\rm s}(z_d)=0.03437$. 
We also need to compute the diameter distance
\begin{equation}
D_{A}(z)=\frac{1}{H_{0}(1+z)}\int_{0}^{z}\frac{dz'}{E(z')}\,.
\label{DAz}
\end{equation}
The dimensionless quantity 
$\bar{D}_{A}=H_{0}D_{A}$ is known 
by integrating the following differential equations:
\begin{eqnarray}
\frac{d\bar{D}_{A}}{dz} & = & -\frac{\bar{D}_{A}}{1+z}+\frac{\sqrt{\bar{\Omega}_{r}}}{(1+z)^{3}}\,,\\
\frac{d\bar{\Omega}_{r}}{dz} & = &
\frac{\bar{\Omega}_r [1-\Omega_{r0}\bar{\Omega}_r
+(3+4s)(1-\Omega_{m0}-\Omega_{r0})
\bar{\Omega}_{r}^{1+s}(1+z)^{-4(1+s)}]}
{(1+z)[1+s(1-\Omega_{m0}-\Omega_{r0})
\bar{\Omega}_{r}^{1+s}(1+z)^{-4(1+s)}]}\,,
\end{eqnarray}
with $\bar{D}_{A}(z=0)=0$.
With the diameter distance, we are now at a place to compute 
the dilation scale \cite{Eisenstein:2005su}
\begin{equation}
D_{V}(z)=\left[ (1+z)^2D_A^2(z)\,z\,H^{-1}(z) \right]^{1/3}=
\left[ D_{A}^{2}(z)\,z\,H_0^{-1} 
\sqrt{\bar{\Omega}_{r}(z)}\right]^{1/3}\,.
\end{equation}
The important observable is the ratio between the sound horizon 
at the drag redshift and the dilation scale
\begin{equation}
\frac{r_{\rm s}(z_{d})}{D_{V}(z)}
=\frac{\bar{r}_{\rm s}(z_d)}
{(\bar{D}_A^2(z)\, z\sqrt{\bar{\Omega}_r})^{1/3}}\,,
\end{equation}
which is dimensionless and does not depend on $H_{0}$. 

We use the BAO data from the 6dFGS \cite{BAO1}, 
SDSS-MGS \cite{BAO2}, 
BOSS \cite{BAO3}, BOSS CMASS \cite{BAO4}, and 
Wiggle Z \cite{BAO5} surveys.
Then, the $\chi^2$ statistics in BAO measurements is given by 
\begin{eqnarray}
\hspace{-0.8cm}
\chi_{\rm BAO}^{2} & = & \frac{1}{0.015^{2}}\left[\frac{r_{\rm s}(z_{d})}{D_{V}(z=0.106)}-0.336\right]^{2}+\frac{1}{\left(\frac{25}{148.69}\right)^{2}}\left[\frac{D_{V}(z=0.15)}{r_{\rm s}(z_{d})}-\frac{664}{148.69}\right]^{2}\nonumber \\
\hspace{-0.8cm}
 &  & +\frac{1}{\left(\frac{25}{149.28}\right)^{2}}\left[\frac{D_{V}(z=0.32)}{r_{\rm s}(z_{d})}-\frac{1264}{149.28}\right]^{2}+\frac{1}{\left(\frac{16}{147.78}\right)^{2}}\left[\frac{D_{V}(z=0.38)}{r_{\rm s}(z_{d})}-\frac{1477}{147.78}\right]^{2}\nonumber \\
 \hspace{-0.8cm}
 &  & +\frac{1}{0.0071^{2}}\left[\frac{r_{\rm s}(z_{d})}{D_{V}(z=0.44)}-0.0916\right]^{2}+\frac{1}{\left(\frac{19}{147.78}\right)^{2}}\left[\frac{D_{V}(z=0.51)}{r_{\rm s}(z_{d})}-\frac{1877}{147.78}\right]^{2}\nonumber \\
 \hspace{-0.8cm}
 &  & +\frac{1}{\left(\frac{20}{149.28}\right)^{2}}\left[\frac{D_{V}(z=0.57)}
 {r_{\rm s}(z_{d})}-\frac{2056}{149.28}\right]^{2}+\frac{1}{0.0034^{2}}
 \left[\frac{r_{\rm s}(z_{d})}{D_{V}(z=0.6)}-0.0726\right]^{2}\nonumber \\
 \hspace{-0.8cm}
 &  & +\frac{1}{\left(\frac{22}{147.78}\right)^{2}}\left[\frac{D_{V}(z=0.61)}{r_{\rm s}(z_{d})}-\frac{2140}{147.78}\right]^{2}+\frac{1}{0.0032^{2}}\left[\frac{r_{\rm s}(z_{d})}{D_{V}(z=0.73)}-0.0592\right]^{2}.
\end{eqnarray}
For the test parameters mentioned above Eq.~(\ref{DAz}), 
we find that  $\chi_{\rm BAO}^{2}=10.443$.

\subsection{SN Ia}
\label{sec:SNI}

The SN Ia can be used as standard candles with known brightness to refer to physical distances.
This is based on the fact that the logarithm of an astronomical object's luminosity seen from a distance of 10 parsecs gives its absolute magnitude $M$, which on the other hand enables us to refer to its brightness. For SN Ia the absolute magnitude at the 
peak of brightness is nearly constant ($M \simeq -19$).
If a supernova at a given redshift $z$ 
is observed with the apparent magnitude $m$, then 
the difference between $m$ and $M$ is related to 
a luminosity distance $d_L(z)$, as
\begin{equation}
\mu (z) \equiv m(z)-M=5 \log \bar{d}_L(z) -5\log h +\mu_0 \,,
\end{equation}
where $\mu_0=42.38$, and 
\begin{equation}
\bar{d}_L(z) \equiv H_0 d_L(z)
=(1+z) \int_0^z \frac{dz'}{E(z')}\,.
\end{equation}
The Hubble expansion rate $H(z)$ is known by 
measuring $m(z)$ for many different redshifts 
($z \lesssim 3$), which allows one to constrain 
dark energy models \cite{Riess,Perl}.

With the distance modulus of our model at hand, 
we can directly compare it with the supernova data and compute 
the $\chi^2$ estimator
\begin{equation}
\chi^2_{\rm SN Ia}=\sum_{i=1}^N
\frac{[\mu_{\rm obs}(z_i)-\mu_{\rm th}(z_i)]^2}
{\sigma_i^2}\,,
\end{equation}
where $N$ is the number of the SN Ia data set, 
$\mu_{\rm obs}(z_i)$ and $\mu_{\rm th}(z_i)$ are 
the observed and theoretical values of the distance 
modulus $\mu(z_i)$, respectively, and $\sigma_i$ are 
the errors on the data. Since the SN Ia data are in the 
low-redshift regime, we can neglect the contribution of  radiation and set $\Omega_r=0$. Furthermore, due to the degeneracy between the absolute magnitude and $h$, 
we will marginalize over $h$. 
For the likelihood analysis, we use the Union 2.1 data sets \cite{Suzuki}.

\subsection{Local measurements of the Hubble expansion rate}

The recent observations of Cepheids in galaxies of SN Ia placed 
the bound on the local value of normalized Hubble constant 
$H_0$, as \cite{Riess:2016jrr}
\begin{equation}
h=0.7324\pm0.0174\,.
\label{H0bound}
\end{equation}
With the knowledge of the comoving sound horizon
it is possible to extract the information of the Hubble 
expansion rate from BAO measurements, as 
\begin{equation}
H(z)\,r_{s}(z_{d})=E(z)\,\bar{r}_{s}(z_{d})
=\frac{(1+z)^2}{\sqrt{\bar{\Omega}_{r}}}
\bar{r}_{s}(z_{d})\,.
\end{equation}
For this purpose we use the recent BOSS data \cite{BAO3} 
in addition to the bound (\ref{H0bound}).
Thus, we define the $\chi^{2}$ statistics associated with the local measurements of $H$ as follows
\begin{eqnarray}
\chi_{H}^{2} & = & \frac{(h-0.7324)^{2}}{0.0174^{2}}+\frac{\left[H(z=0.38)\,r_{s}(z_{d})-81.5\times147.78/299792.458\right]^{2}}{(2.6\times147.78/299792.458)^{2}}\nonumber \\
 &  & +\frac{\left[H(z=0.51)\,r_{s}(z_{d})-90.5\times147.78/299792.458\right]^{2}}{(2.6\times147.78/299792.458)^{2}}\nonumber \\
 &  & +\frac{\left[H(z=0.61)\,r_{s}(z_{d})-97.3\times147.78/299792.458\right]^{2}}{(2.9\times147.78/299792.458)^{2}}\,.
\end{eqnarray}
For the test parameters $\Omega_{m0}=0.3$, $h=0.6774$, and $s=0.1$, we obtain $\chi_{H}^{2}=30.43$.

\subsection{RSD}

For the perturbations relevant to the RSD measurements 
(sub-horizon modes with $k \gg aH$) it was shown in 
Ref.~\cite{DeFelice:2016uil}
that the quasi-static approximation is sufficiently accurate to 
describe the evolution of the matter density contrast 
$\delta$, so we resort to Eq.~(\ref{deleq2}) without taking 
into account the radiation ($\Omega_r=0$). 
In terms of the derivative with respect to ${\cal N}=\ln a$, 
we can rewrite Eq.~(\ref{deleq2}) in the form 
\begin{equation}
\delta''+\frac{1-3w_{\rm eff}}{2}\delta'
-\frac{4\pi G_{\rm eff} \rho_m}{H^2} \delta =0\,,
\label{delsub}
\end{equation}
where $w_{\rm eff}=-1-2\dot{H}/(3H^2)$ is the effective 
equation of state whose explicit form in our model 
is given by Eq.~(\ref{weff}). 
On using the matter density parameter 
$\Omega_m=8\pi G \rho_m/(3H^2)$, where 
$G=1/(8\pi M_{\rm pl}^2)$ is the bare gravitational 
constant, Eq.~(\ref{delsub}) reduces to 
\begin{equation}
\delta''+\frac{1+(3+4s)\Omega_{{\rm DE}}}{2(1+s\Omega_{{\rm DE}})}
\delta'-\frac{3}{2}\,\frac{G_{{\rm eff}}}{G}\,
(1-\Omega_{{\rm DE}})\delta=0\,.
\label{delsub2}
\end{equation}
For the integration of this differential equation, 
we use the effective gravitational coupling given by 
Eq.~(\ref{Geff2}).

The $\Lambda$CDM model corresponds to $s=0$ and 
$G_{\rm eff}=G$ in Eq.~(\ref{delsub2}), in which 
case there is the growing-mode solution 
$\delta \propto a$ during the matter dominance 
($\Omega_{\rm DE} \simeq 0$). 
In our model $\Omega_{\rm DE} \simeq 0$ 
and $G_{\rm eff} \simeq G$ in the early matter era, 
so the evolution of $\delta$ at high redshifts ($z \gg 1$) 
is very similar to that in the 
$\Lambda$CDM model. 
Hence we choose the initial conditions satisfying 
$\delta=\delta'$ in the deep matter era.
The difference from the $\Lambda$CDM model arises 
at low redshifts where $\Omega_{\rm DE}$ and 
$G_{\rm eff}$ deviate from 0 and $G$, respectively.

We define $\sigma_8$ as the amplitude of over-density 
at the comoving $8\,h^{-1}$ Mpc scale.
To compute the $\chi^2$ estimator of RSD measurements, 
we introduce the following quantity
\begin{equation}
y(z) \equiv f(z) \sigma_8(z)\,,\qquad f(z) 
\equiv \frac{\delta'}{\delta}\,.
\end{equation}
Since the behavior of perturbations in our model is very close to that in the $\Lambda$CDM model at high redshifts, the two models should have similar initial conditions 
of $\delta$. This implies that
$\sigma_8^{\rm Proca}(z_i)\approx\sigma_8^{\Lambda
{\rm CDM}}(z_i)$ at an initial redshift $z=z_i\gg1$.
Since we assume that the radiation is negligible, we choose such an initial redshift
to be in the deep matter era, that is, at $N_i=-6$
(around $z_i\simeq400$).

For the $f \sigma_8$ data extracted from RSD measurements, 
we use those listed in Table I of 
Ref.~\cite{DeFelice:2016ufg}. 
This includes the recent data of FastSound 
at the redshift $z=1.36$ \cite{Okumura:2015lvp}. 
We define the $\chi^2$ estimator, as 
\begin{equation}
\chi^2_{\rm RSD}=\sum_{i=1}^N
\frac{[y_{\rm obs}(z_i)-y_{\rm th}(z_i)]^2}
{\sigma_i^2}\,,
\end{equation}
where $N$ is the number of the RSD data, 
$y_{\rm obs}(z_i)$ and $y_{\rm th}(z_i)$
are the observed and theoretical values of $y(z_i)$ 
at redshift $z_i$ respectively, and 
$\sigma_i$'s are the errors on the data. 

Among other measurements discussed in this section,
the RSD data are only those strictly connected with the 
growth of perturbations. Since the effective gravitational 
coupling (\ref{Geff2}) depends on many model parameters 
like $q_V$ and $c_S^2$, there should be some level of degeneracy of model parameters compared to the analysis 
based on the background expansion history.
Reflecting this situation, we will focus on the case 
$\beta_5=0$ for the likelihood analysis including the RSD data.

\section{Background constraints} 
\label{backconsec}

As a first analysis of the theory, we study how the background cosmic 
expansion history can fit the observational data sets of 
CMB, BAO, SN Ia, and the Hubble expansion rate. 
In this section we do not take into account the RSD data,
as they directly deal with the growth of matter perturbations. 
Then, we focus on the likelihood analysis 
for the following parameters:
\begin{equation}
\Omega_{m0},\quad h, \quad s\, .
\label{backpara}
\end{equation}
The analysis of the background alone is simpler than the one including 
the perturbations, as the space of parameters reduces to 
a three-dimensional one. 
It should be noticed that, compared to the $\Lambda$CDM model, 
our model has only one additional background parameter, $s$. 
Furthermore, the $\Lambda$CDM model corresponds to $s=0$ 
at the background level, so the direct comparison 
between the two models is straightforward.

For the Monte-Carlo-Markov-Chain (MCMC) sampling, we need to 
put some priors on the allowed parameter space of the three parameters. 
We will choose sensible priors for the
parameters (\ref{backpara}) as follows: 
\begin{itemize}
\item For the density parameter of non-relativistic matter, 
$0.1\leq \Omega_{m0} \leq 0.5 $.
\item For the normalized Hubble constant, $0.6 \leq h \leq 0.8$.
\item For the deviation parameter $s$ from the $\Lambda$CDM model, 
$-0.3\leq s \leq 0.8$.
\end{itemize}
Since the analysis including perturbations is not performed in this section, 
we do not put priors on the quantities associated with perturbations 
(e.g., $q_V>0$, $c_S^2>0$ etc).
As we will see in Sec.~\ref{perconsec}, some of the
results can change by taking into account perturbations and 
the allowed parameter space may be reduced. Nonetheless, 
it is worthy of investigating first how our model can fit the data 
at the background level compared to 
the $\Lambda$CDM model.
We perform the MCMC sampling over the allowed 
three-dimensional parameter space and compute 
\be
\chi_{\rm back}^2=\chi^2_{\rm CMB}+\chi^2_{\rm BAO}
+\chi^2_{\rm SN Ia}+\chi^2_H\,.
\ee
The best fit corresponds to the case in which 
$\chi_{\rm back}^2$ is minimized.

\begin{figure}
\begin{center}
\includegraphics[height=6.5in,width=6.5in]{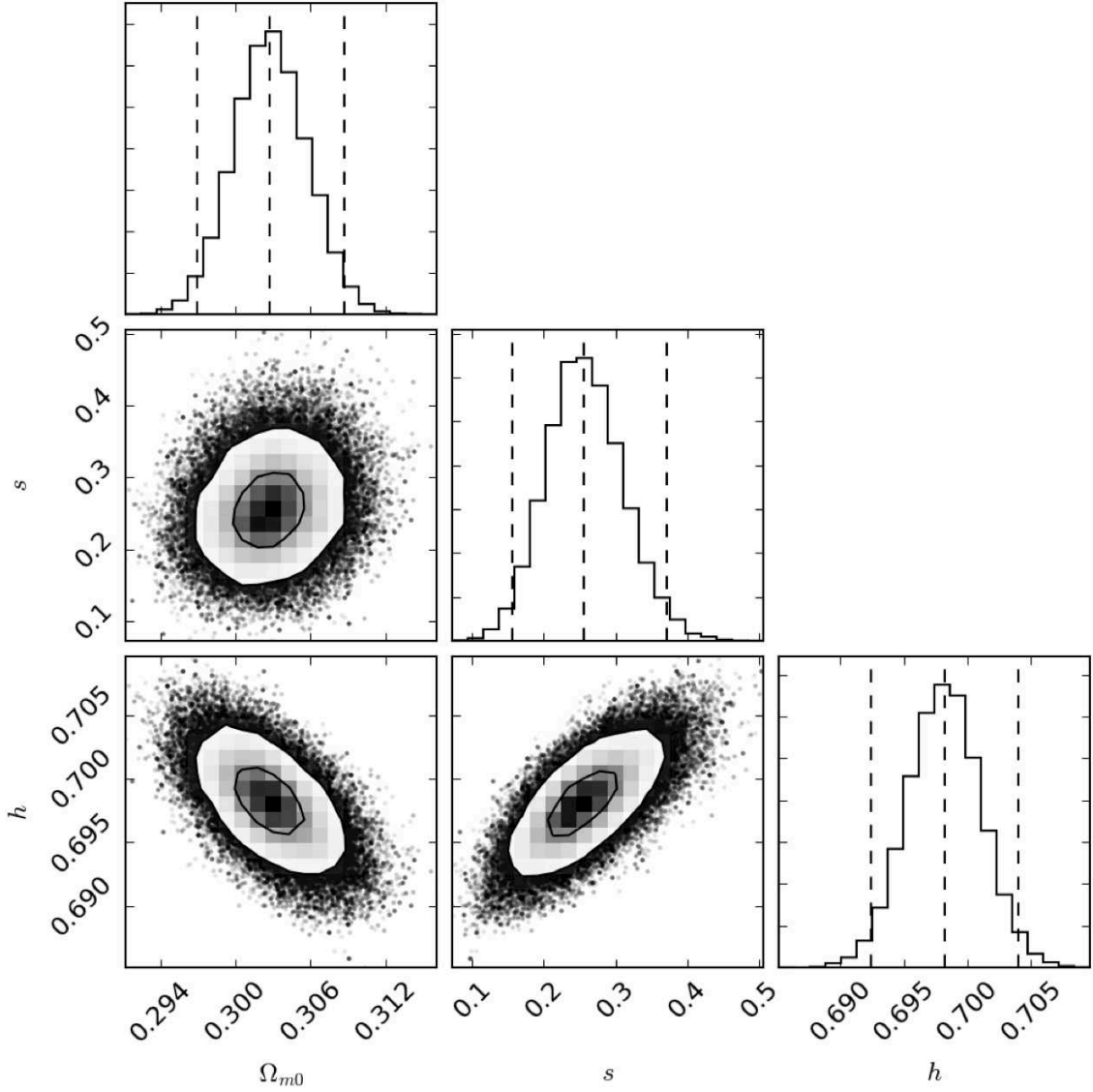}
\end{center}
\caption{\label{fig1}
Observational bounds on the three parameters $\Omega_{m0},s,h$  
constrained by the data of the CMB, BAO, SN Ia, and the Hubble 
expansion rate. The RSD data are not taken into account in the analysis.
Since the analysis is based on the background cosmic expansion 
history, we do not consider the conditions associated with ghosts 
and stabilities of perturbations. 
{}From top to bottom, the right panels in each column are 
one-dimensional probability distributions of the parameters 
$\Omega_{m0},s,h$, respectively. The vertical dashed lines 
correspond to the best fit (central) and the 2$\sigma$ confidence 
limits (outside). The other panels are the two-dimensional likelihood 
contours in the $(s,\Omega_{m0})$, $(h, \Omega_{m0})$, 
and $(h,s)$ planes with $1\sigma$ (inside) and $2\sigma$ (outside) boundaries.
The $\Lambda$CDM model, which corresponds to $s=0$, 
is disfavored over the model with $s>0$ from the background analysis.}
\end{figure}

In Fig.~\ref{fig1} we plot one-dimensional probability distributions of the 
parameters (\ref{backpara}) and two-dimensional observational contours for the 
combination of these three parameters.
The one-dimensional probability distributions show that 
the minimum value of $\chi_{\rm back}^2$ does exist for the following (approximated) values:
\be
\Omega_{m0}=0.3027\,,\qquad
h=0.6981\,,\qquad
s=0.254\,,
\label{Omeback}
\ee
for which 
\be
\chi^2_{\rm back,min}\approx590.4\,.
\ee
The existence of a minimum around $s=0.25$ shows that 
the model with $s>0$ is favored over the $\Lambda$CDM model 
at the background level.
The 2$\sigma$ constraints on the three parameters are given by
\begin{eqnarray}
\Omega_{m0} &=& 0.3027^{{}+0.0060}_{{}-0.0057}\,,
\label{OmemconB}\\
h &=& 0.6981^{{}+0.0059}_{{}-0.0057}\,,
\label{hconB}\\
s &=& 0.254^{{}+ 0.118}_{{}-0.097}\,.
\end{eqnarray}
This means that the $\Lambda$CDM model is disfavored 
over the model with $s>0$ even at the 2$\sigma$ level. 
We note that the extended scalar Galileon model \cite{exscaGa1} 
has a tracker solution whose background evolution is the same as that in the 
model under consideration.
In Ref.~\cite{exscaGa2} two of the present authors 
performed the likelihood analysis by using the data of 
the CMB (WMAP7), BAO, SN Ia, and derived the bound 
$s=0.034^{+0.327}_{-0.034}$ (95\,\% CL). 
With the new data of the CMB (Planck), BAO, SN Ia, and the Hubble expansion 
rate, the constraint on $s$ is shifted toward larger values.

In the $\Lambda$CDM model, the best-fit values of $\Omega_{m0}$ and 
$h$ constrained by the Planck CMB data are around 
$\Omega_{m0} \approx 0.31$ and $h \approx 0.68$, 
respectively \cite{Planck2015}. 
These best-fit values are in tension with their 
low-redshift measurements, which generally favor lower 
$\Omega_{m0}$ and higher $h$, see, e.g., Eq.~(\ref{H0bound}). 
The model with $s>0$ can reduce such a tension with the shift 
toward smaller $\Omega_{m0}$ and larger $h$.
We can confirm this property in the probability distributions 
of $\Omega_{m0}$ and $h$ in Fig.~\ref{fig1}.
 
\section{Observational constraints including the RSD data} 
\label{perconsec}

If we take into account the evolution of matter perturbations, the 
likelihood results can be subject to change in two different ways: 
1) the stability conditions of perturbations, which need to hold 
at all times, generally reduce the allowed parameter space; and 
2) the RSD contribution can shift observational bounds of 
model parameters. 
We set the following additional priors to those 
used in Sec.~\ref{backconsec}:
\begin{itemize}
\item The no-ghost conditions for scalar, vector, and tensor 
perturbations to apply at all times, 
i.e., $q_T>0,q_V>0,Q_S>0$.
\item The stability conditions associated with the propagation speeds 
of scalar, vector, and tensor perturbations, 
i.e., $c_T^2>0,c_V^2>0,c_S^2>0$. 
We also put the priors that $c_T^2, c_V^2, c_S^2$ are initially 
smaller than $10^3$ to avoid the divergences of these quantities 
in the asymptotic past.
\item The condition $c_T^2>1$ to be valid {\it today}. 
This is for evading the Cherenkov radiation bound 
$1-c_T<2 \times 10^{-15}$ today \cite{Cheren1,Cheren2}.
\item $0<p\leq 25$ for keeping the parameter $p$ 
positive and of order unity.
\item $-10^3\leq\beta\leq-10^{-9}$, $10^{-13}\leq\lambda\leq15$.
The reason for the choices of negative $\beta$ and positive $\lambda$ 
were explained in Sec.~\ref{conmodelsec}.
We choose flat distributions for the natural logarithms of these variables.
\item $-4\leq\ln(q_{V})\leq15$. 
We choose the values of $q_{V}$ to be not very close to 0 
to avoid the strong coupling problem.
\end{itemize}
For simplicity we set the parameter $\beta_5$ to zero, as keeping it 
non-zero in the analysis does not change the final results
significantly. The likelihood results seem to be flat in this direction, 
such that the observational data do not notably constrain this parameter.

We perform a MCMC analysis and compute
\be
\chi^2=\chi^2_{\rm CMB}+\chi^2_{\rm BAO}
+\chi^2_{\rm SN Ia}+\chi^2_H
+\chi^2_{\rm RSD}\,,
\ee
in order to quantify how the RSD 
data affect the available parameter space constrained at the 
background level. 
We verify that the role of no-ghost and stability conditions is
important, because some of the parameter space preferred by
the background analysis (performed in Sec.~\ref{backconsec}) 
does not possess a stable cosmological evolution. 
For example, the best fit obtained by the background analysis 
does not in general satisfy no-ghost conditions of perturbations.
Furthermore we find that the RSD data affect constraints on some parameters like $p$, 
but they only give mild bounds on other parameters such as $\beta, \lambda, q_V$. Nonetheless, the RSD data also contribute to shifting/reducing the parameter space for the background parameters $\Omega_{m0},h,s$, because the matter perturbation equation depends on those parameters.

The MCMC likelihood analysis shows that there is a quite large 
degeneracy in terms of the minimum $\chi^2$, i.e., several different 
parameters associated with perturbations can lead to similar values of $\chi^2$. 
In other words, the model does not seem to have one unique global minimum of 
$\chi^2$, but there are several of those minima for different sets of parameters. 
We can pick up one example of such a low value of $\chi^2$. 
For instance, we obtain the minimum value 
\be
\chi^2_{\rm min}=625.6\,,
\label{chismin}
\ee
for the following (approximated) values:
\begin{eqnarray}
& &
\Omega_{m0}=0.299\,,\qquad
h=0.6962\,,\qquad
s=0.16\,,\nonumber \\
& &
p=2.69\,,\qquad
\ln(-\beta)= 0.33\,,\qquad
\ln q_V=-1.731\,,\qquad
\ln \lambda=-16.9\,.
\end{eqnarray}
The corresponding 2$\sigma$ bounds on these 
parameters are given, respectively, by 
\begin{eqnarray}
& &
\Omega_{m0}=0.299^{{+}0.006}_{{}-0.006}\,,
\label{Omemcon} \\
& &
h=0.696^{{+}0.006}_{{-}0.005}\,,
\label{hcon}\\
& &
s=0.16^{{+}0.08}_{{-}0.08}\,,\label{scon}\\
& &
p=2.69^{{+}19.91}_{{-}0.73}\,,\\
& &
\beta=-1.39^{{+}0.75}_{{-}492.28} \,,\\
& &
{\bar q}_V \leq q_V < 164 \,,\\
& &
{\bar \lambda} \leq \lambda < 0.015\,,
\end{eqnarray}
where ${\bar q}_V$ and ${\bar \lambda}$ are the lower limits from the assumed prior/numerical precision.

\begin{figure}
\begin{center}
\includegraphics[height=6.5in,width=6.5in]{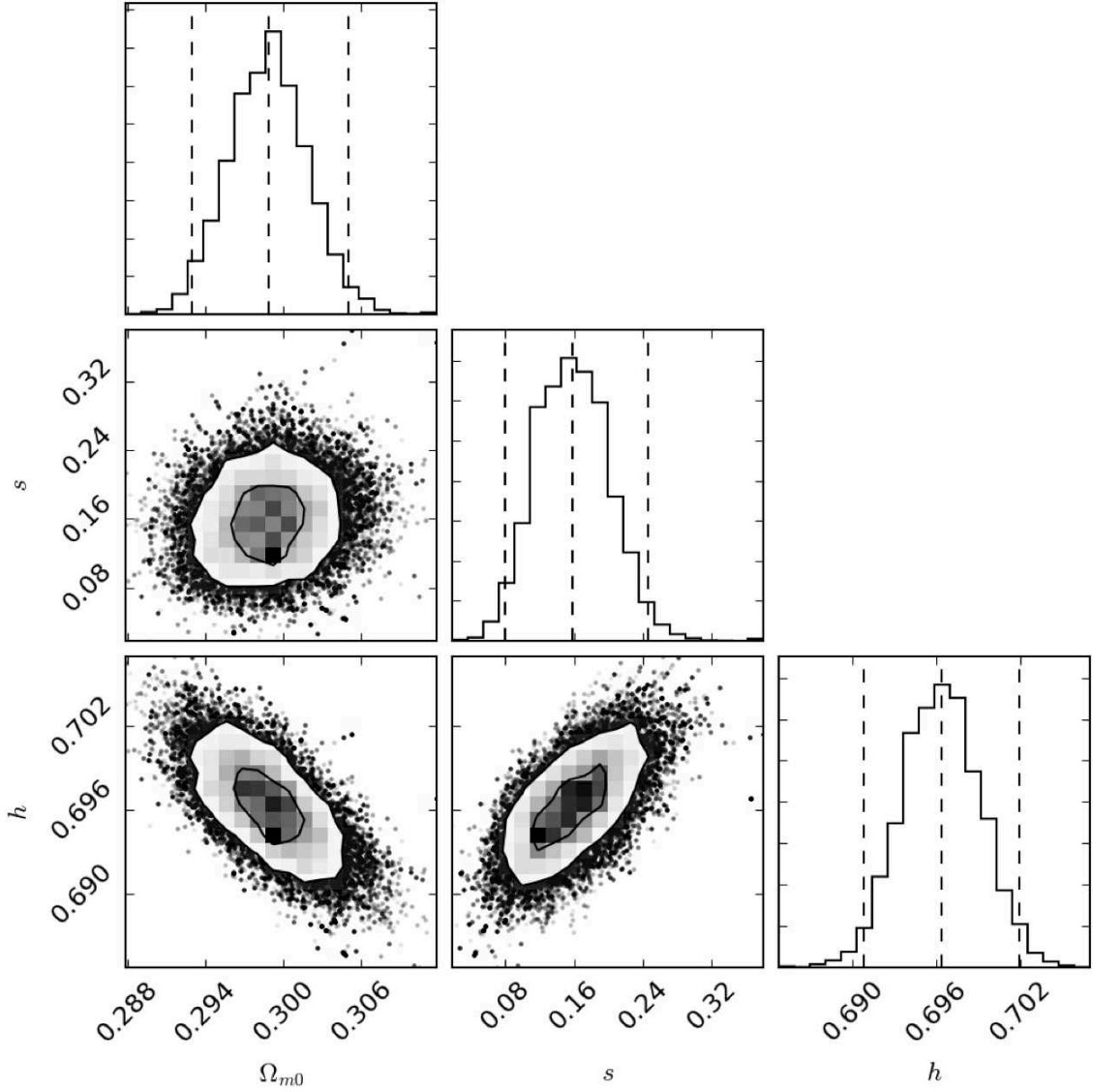}
\end{center}
\caption{\label{fig2}
Observational bounds on the three parameters 
$\Omega_{m0},s,h$ derived by adding the RSD data 
to the data of the CMB, BAO, SN Ia, and the Hubble expansion rate.
The no-ghost and stability conditions are also taken 
into account. The meanings of one-dimensional probability 
distributions and two-dimensional likelihood contours are 
the same as those explained in the caption of Fig.~\ref{fig1}.
The $\Lambda$CDM model is still disfavored 
over the model with $s>0$.}
\end{figure}

\begin{figure}
\begin{center}
\includegraphics[height=6.5in,width=6.5in]{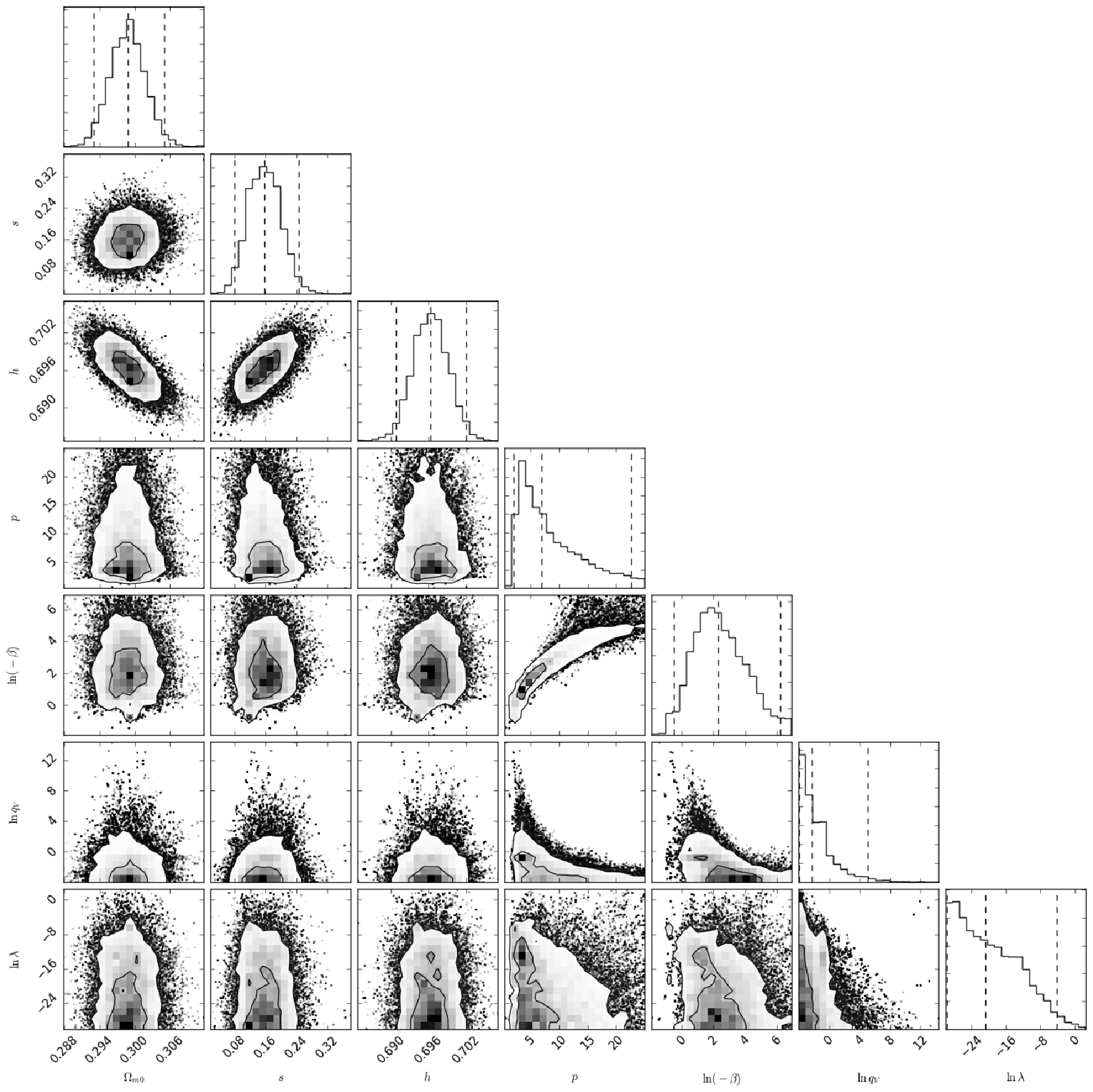}
\end{center}
\caption{\label{fig3}
Observational constraints on the seven model parameters 
$\Omega_{m0},s,h,p,\beta,q_V,\lambda$ derived by the 
joint data analysis of the CMB, BAO, SN Ia, the Hubble expansion rate, 
and RSD data, with no-ghost and stability conditions taken 
into account. The meanings of one-dimensional probability 
distributions and two-dimensional likelihood contours are 
the same as those explained in the caption of Fig.~\ref{fig1}. 
The three parameters $\Omega_{m0},s,h$ associated with the 
background are tightly constrained, but the bounds on the 
four parameters $p, \beta, q_V, \lambda$ are quite weak.}
\end{figure}

The probability distributions and the likelihood contours for 
the background parameters $\Omega_{m0},s,h$ and the 
full parameters $\Omega_{m0},s,h,p,\beta,q_V,\lambda$ are 
plotted in Figs.~\ref{fig2} and \ref{fig3}, respectively.
The RSD data together with the stability conditions 
influence the parameter space in several ways. 
We summarize the main results in the following.
\begin{itemize}
\item 
Inclusion of the RSD data tends to reduce the best-fit 
value of $s$ (compared to $s \simeq 0.26$ constrained from 
the background), but still a positive value of $s$ around 0.16 
is favored, see Fig.~\ref{fig2}.
This implies that the RSD data generally prefer 
the model with lower $s$. Indeed, the RSD data alone 
can be consistent with the case $s=0$.
This is mostly attributed to the fact that the models with 
a larger $s$ tend to give rise to a larger effective gravitational 
coupling with $G_{\rm eff}>G$.
There is a tendency that the current RSD measurements favor 
the cosmic growth rate smaller than that predicted by 
GR \cite{Ade:2015rim}. In generalized Proca theories 
it is possible to realize $G_{\rm eff}<G$, but this occurs 
at the expense of choosing the value of $q_V$ 
close to 0 \cite{DeFelice:2016uil}. 
To avoid the strong coupling problem of vector perturbations 
we set the prior $q_V \lesssim 10^{-2}$, in which case 
$G_{\rm eff}$ cannot be significantly smaller than $G$. 
Since the effect of weak gravity arising from small $q_V$ 
is limited, the modification of $G_{\rm eff}$ induced by 
the change of $s$ tends to be more important.
In Fig.~\ref{fig3} we observe that the parameter $q_V$ is 
loosely constrained.

\item 
The data other than RSD favor a non-zero positive value 
of $s$. Therefore, combining all these different contributions, 
we obtain the bound (\ref{scon}), 
which is smaller than $s=0.25$.
On performing a MCMC sampling for the $\Lambda$CDM 
model with the two parameters $\Omega_{m0}$ and $h$, 
we find that the best-fit case corresponds to 
$\chi^2_{\Lambda{\rm CDM}}=642.7$ with
$\Omega_{m0}=0.298$ and $h=0.688$.
Since this value of $\chi^2$ is larger than (\ref{chismin}), 
the model with $s \approx 0.16$ can fit the joint 
observational constraints of the CMB, BAO, SN Ia, 
the Hubble expansion rate, and RSD better 
than the $\Lambda$CDM model.

\item Comparing the bounds (\ref{Omemcon}) and (\ref{hcon}) with 
Eqs.~(\ref{OmemconB}) and (\ref{hconB}), the constraints on 
$\Omega_{m0}$ and $h$ derived by adding the RSD 
data to the data associated with the background expansion history 
are not subject to significant modifications relative to those obtained 
without the RSD data.

\item Since only the RSD data are affected by the four parameters $p,\beta,q_V, \lambda$ associated with perturbations, we find that the constraints on them are mild and that some degeneracy of $\chi^2$ exists among different model parameters. We expect that this degeneracy may be reduced by including 
other independent observational data relevant to perturbations. 
\end{itemize}
%

\section{Conclusions} 
\label{consec}

The recently proposed generalized Proca interactions constitute promising alternative theories of gravity on large scales. These derivative vector-tensor interactions are constructed in such a way that the resulting theories contain only five propagating degrees of freedom, three of them originating from the massive vector field \cite{Heisenberg:2014rta,Allys:2015sht,Jimenez:2016isa}. They establish a consistent framework for the late-time cosmic acceleration. 
On the FLRW background, the temporal component of the vector field gives rise to interesting de Sitter solutions relevant to dark energy. Even if the temporal component is not dynamical, its auxiliary role results in promising de Sitter attractors as it was shown in Ref.~\cite{DeFelice:2016yws}. 

In this work, we have placed observational constraints on a class of dark energy models in the framework of generalized Proca theories. We have first summarized the key findings of the background evolution and 
stability analysis of perturbations performed in 
Refs.~\cite{DeFelice:2016yws,DeFelice:2016uil}. 
The background dynamics is rather simple and dictated by the three 
parameters $\Omega_{m0},h,s$, where $s$ represents the deviation from 
the $\Lambda$CDM model. For the evolution of matter perturbations, 
we have used the equation derived under the quasi-static approximation
on sub-horizon scales, whose validity was explicitly checked in 
Ref.~\cite{DeFelice:2016uil}. We have also taken into account 
conditions for avoiding ghosts and Laplacian instabilities of 
tensor, vector, and scalar perturbations. 
The perturbations carry four additional parameters 
$p,\beta,q_V,\lambda$ than those associated with the background.

At the background level, we have exploited the data sets of 
CMB distance priors, BAO, SN Ia (Union 2.1), and 
local measurements of the Hubble expansion rate.
We have found that the MCMC analysis constrains the parameter $s$ 
to be $s=0.254^{+0.118}_{-0.097}$ (95\,\%\,CL) from the background 
cosmic expansion history, so the model with $s>0$ can have a good fit to the data compared to the $\Lambda$CDM model.
This is attributed to the fact that existence of the additional 
parameter $s$ can reduce the tensions of the parameters 
$\Omega_{m0}$ and $h$ between early-time and 
late-time data sets.

Including the RSD data as well as no-ghost and stability conditions 
of perturbations, the bound on the parameter $s$ is shifted to 
$s=0.16^{+0.08}_{-0.08}$ (95\,\%\,CL).
This shift is mostly related to the fact that the RSD data tend to 
favor lower values of $s$ for realizing $G_{\rm eff}$ close to $G$. 
Existence of the intrinsic vector mode can lead to a 
$G_{\rm eff}$ 
smaller than $G$ for $q_V$ close to 0, but this effect is limited 
by the fact that the vector perturbation has a strong coupling problem 
for such small values of $q_V$. 
Since the data other than RSD prefer positive values of $s$ away from 0, 
the joint data analysis including both the background and the RSD data 
still favor the model with $s>0$ over the $\Lambda$CDM model. 
We have also found that, expect for the background quantities 
$\Omega_{m0},h,s$, the observational bounds on other parameters 
are not stringent.

As we have seen in this work, the derivative interactions 
in generalized Proca theories facilitate the background evolutions 
quite generically, which give rise to the dynamics for alleviating the tension 
between early-time and late-time data sets. 
It is still possible that the tension present in the $\Lambda$CDM model 
may be related to some systematic errors in one (or more) data sets. 
Nonetheless, we find it interesting at least to have shown that the 
cosmological background fitting the data well can naturally 
follow from generalized Proca theories.

We note that beyond-generalized Proca theories recently 
proposed in Ref.~\cite{Heisenberg:2016eld} share the
same background evolution as the model (\ref{G2345}).
Nevertheless, the presence of additional terms yields distinctive 
features at the level of perturbations. For instance, it is possible to weaken the gravitational coupling with non-relativistic matter further, 
e.g., $G_{\rm eff} \approx 0.8G$ today \cite{Nakamura:2017dnf}. 
The smaller effective gravitational coupling, which may fit the RSD data 
better than the model studied in this paper, arise from 
beyond-generalized Proca interactions rather than from 
intrinsic vector modes with $q_V$ close to 0. 
It will be of interest to place observational 
constraints on such models as well by adding other data associated 
with perturbations, e.g., the integrated Sachs-Wolfe effect of CMB 
and weak lensing. We hope that future high-precision observations 
will allow us to distinguish the models in the framework of 
(beyond-)generalized Proca theories from the $\Lambda$CDM 
model further.

\begin{acknowledgments}
ADF is supported by the JSPS KAKENHI Grant Numbers 16K05348 
and 16H01099. LH acknowledges financial support from Dr.\ Max R\"ossler, the Walter Haefner 
Foundation and the ETH Zurich Foundation.
ST is supported by the Grant-in-Aid for Scientific Research Fund of the JSPS No.~16K05359 and a MEXT KAKENHI Grant-in-Aid for Scientific Research on Innovative Areas 
``Cosmic Acceleration'' (No.\,15H05890).
\end{acknowledgments}


\end{document}